\documentclass[a4paper,11pt]{article}

\usepackage{jheppub} 

\usepackage[T1]{fontenc} 
\usepackage{comment}
\usepackage{pdflscape}

\usepackage{pgfplots}
\pgfplotsset{width=10cm,compat=1.9}
\usepgfplotslibrary{fillbetween}
\newcommand{\be}{\begin{equation}}
\newcommand{\ee}{\end{equation}}

\newcommand{\bea}{\begin{eqnarray}}
\newcommand{\eea}{\end{eqnarray}}
\numberwithin{equation}{section}

\definecolor{darkgreen}{rgb}{0, 0.6, 0}

\newcommand{\V}{{\cal V}}

\def\rmi{{\rm i}}
\title{\boldmath Anisotropic Dark Energy from String Compactifications}

\author[a]{Diego Gallego}
\author[b]{J. Bayron Orjuela-Quintana,}
\author[b]{and C\'esar A. Valenzuela-Toledo}

\affiliation[a]{Grupo de F\'isica de Altas Energ\'ias\\
	Escuela de F\'isica, Universidad Pedag\'ogica y Tecnol\'ogica de Colombia UPTC,\\ Avenida Central de Norte, Tunja, Colombia.}
\affiliation[b]{Departamento de F\'isica, Universidad del Valle, Ciudad Universitaria Mel\'endez, 760032, Cali, Colombia}

\emailAdd{diego.gallego@uptc.edu.co}
\emailAdd{john.orjuela@correounivalle.edu.co}
\emailAdd{cesar.valenzuela@correounivalle.edu.co}

\abstract{We explore the cosmological dynamics of a minimalistic yet generic string-inspired model for multifield dark energy. Adopting a supergravity four-dimensional viewpoint, we motivate the model's structure arising from superstring compactifications involving a chiral superfield and a pure $U(1)$ gauge sector. The chiral sector gives rise to a pair of scalar fields, such as the axio-dilaton, which are kinetically coupled. However, the scalar potential depends on only one of them, further entwined with the vector field through the gauge kinetic function. The model has two anisotropic attractor solutions that, despite a steep potential and thanks to multifield dynamics, could explain the current accelerated expansion of the Universe while satisfying observational constraints on the late-times cosmological anisotropy. Nevertheless, justifying the parameter space allowing for slow roll dynamics together with the correct cosmological parameters, would be challenging within the landscape of string theory. Intriguingly, we find that the vector field, particularly at one of the studied fixed points, plays a crucial role in enabling geodesic trajectories in the scalar field space while realizing slow-roll dynamics with a steep potential. This observation opens a new avenue for exploring multifield dark energy models within the superstring landscape.
}

\keywords{Cosmology of theories beyond the SM, String and Brane Phenomenology, Superstring Vacua, Dark energy, Multi-field dark energy, Swampland}

\begin{document} 
\maketitle
\flushbottom

\section{Introduction}
\label{Sec: Introduction}

For over two decades, there has been strong evidence of a present accelerated epoch in the universe \cite{SupernovaSearchTeam:1998fmf, SupernovaCosmologyProject:1998vns}, supported by recent observations of the Cosmic Microwave Background (CMB) \cite{WMAP:2012fli, Komatsu:2014ioa, Planck:2018vyg} (also see \cite{Agatsuma:2022ewd}). The simplest explanation for this behaviour is a vacuum energy or cosmological constant $\Lambda$ \cite{Peebles:2002gy, Padmanabhan:2002ji}, which is accommodated in the so-called $\Lambda$CDM (Cold Dark Matter) or concordance model. Interestingly, this model started to gain traction as a strong candidate for describing cosmological data even before such observations were made \cite{Ostriker:1995rn}.

There are several reasons to explore alternative approaches. Firstly, the vacuum energy interpretation suffers from the lack of a solid theoretical prediction, with a notable concern being the discrepancy of $120$ orders of magnitude~\cite{Martin:2012bt}. Secondly, recent results have led to cosmological tensions that cannot, at least easily, be adequately accommodated within this framework~\cite{Perivolaropoulos:2021jda}. Among these challenges, the most significant is the Hubble constant discrepancy \cite{Riess:2016jrr, Riess:2020fzl, Verde:2019ivm}. (For a comprehensive overview and the current status, please refer to \cite{Schoneberg:2021qvd}.) Furthermore, exploring alternative models enables us to address issues such as the coincidence problem. All this is happening against the backdrop of the dawn of precision cosmological experiments (e.g., Refs.~\cite{DESI:2016fyo, Amendola:2016saw, Gebhardt:2021vfo}), which hold the promise of refining our understanding of the early and late universe.

The goal is to replicate an accelerating late phase of cosmological evolution by emulating the effects of the cosmological constant through field dynamics \cite{ratraCosmologicalConsequencesRolling1988,Wetterich:1994bg}. This strategy aligns with the concept of Dark Energy (DE) and seeks to elucidate the origin of the fields and their couplings, potentially arising from an ultraviolet completion like string theory. In fact, in a broader context, any ultraviolet completion will extend this foundational model by introducing additional fields, consequently leading to non-standard cosmologies.

Despite the potential conceptual challenges surrounding its compatibility with accelerated universes \cite{Hellerman:2001yi, Fischler:2001yj}, superstring theory provides a top-down approach that serves as both a playground for model construction and a guiding principle in ensuring consistency within possible ultraviolet completions that incorporate quantum gravity. Remarkably, the phenomenology of string theory has made significant strides over recent decades, offering a wealth of insights into potential fields and their interactions. Although precise and exact outcomes remain elusive due to the intricacies of transitioning from the 10-dimensional theory to the 4-dimensional effective description, guidelines have surfaced, exemplified by the \emph{swampland conjectures} \cite{Vafa:2005ui, Ooguri:2006in, Ooguri:2016pdq,Brennan:2017rbf,Palti:2019pca,VanRiet:2023pnx}. These conjectures shed light on what can be anticipated from theories aiming for a coherent ultraviolet completion that encompasses gravity.\footnote{The pursuit of incorporating any model within the framework of string theory is more than an academic exercise; it emerges as an essential endeavour. This arises from the likelihood that explaining an accelerated epoch necessitates substantial contributions from a microscopic description that goes beyond the effective field theory approximation for point particles \cite{Brandenberger:2022sga}.}

Scalar fields are pervasive in effective 4-dimensional theories derived from string theory. They describe various couplings (e.g., dilaton modulus), internal dimensions' size and shape (e.g., Kähler modulus), brane locations, and more. Thus, the simplest scenarios involve single scalar field models. Encouragingly, several positive results have already been achieved. For instance, there are potential UV descriptions of generalizations of the Chaplygin gas derived from D-branes \cite{Bergner:2001db, Hassaine:2001is,Bento:2002ps}, stringy axions \cite{Panda:2010uq, Cicoli:2012tz, DAmico:2018mnx}, and early dark energy scenarios like the one presented in Ref.~\cite{Cicoli:2023qri}.

Additionally, there are multifield scenarios, as seen in references such as \cite{Kaloper:2005aj, Kaloper:2008qs, Blaback:2013fca, Kamionkowski:2014zda, Alexander:2019rsc, McDonough:2022pku}. In many cases, multifield scenarios address issues present in the single field case and include elements like the axionic superpartner for scalar fields. These multifield situations are not only more natural due to the expectation of many scalar fields in UV completions but also align with the \emph{swampland conjectures}. These conjectures demand very steep scalar potentials, often in conflict with working scenarios of quintessence. Interestingly, a non-trivial curvature in the scalar field manifold can evade such constraints, as shown in Ref.~\cite{Achucarro:2018vey} (see also Ref.~\cite{Akrami:2020zfz}), and permit behaviours resembling inflation \cite{Brown:2017osf, Cicoli:2012tz, Christodoulidis:2019jsx, Alexander:2019rsc, Bernardo:2022ztc}. An interesting situation of multifield dynamics can simultaneously address the puzzle of dark matter, as shown for example in Ref.~\cite{Gasperini:2001pc} and, recently in Ref.~\cite{vandeBruck:2019vzd,Bernardo:2022ztc, Gomes:2023dat}. The former work builds on a string inspired model of the dilaton that couples with a hidden sector, but does no regard its superpartner. This is done in Ref.~\cite{sonnerRecurrentAccelerationDilatonaxion2006} (also in Ref.~\cite{vandebruckQuintessenceDynamicsTwo2009} with more involved kinetic mixings) where, however, no any other energy components are included. This system, together with a barotropic perfect fluid is instead studied in Refs.~\cite{Cicoli:2020cfj,cicoliOutSwamplandMultifield2020} and with a direct universal coupling in Ref.~\cite{catenaAxionDilatonCosmology2008}.

Vector fields are, indeed, a more intrinsic component of nature than scalar fields and are anticipated to be inherent in any extension of the standard model. The issue of consistency becomes less straightforward in this context, owing to the potential emergence of ghost fields. Nevertheless, substantial progress has been made in comprehending a bottom-up approach for constructing effective Lagrangians involving these fields, as discussed in references such as \cite{Heisenberg:2014rta,Tasinato:2014eka,BeltranJimenez:2016rff,Allys:2015sht,Allys:2016jaq,DeFelice:2016yws,Heisenberg:2018mxx}, facilitating a broad spectrum of cosmological investigations.

In the line of the present study, dynamical dark energy scenarios have been formulated, wherein vector fields play a pivotal role. For example, Refs.~\cite{Rinaldi:2015iza, Alvarez:2019ues, Guarnizo:2020pkj} introduces vector fields associated with an $SO(3)$ symmetry, forming what is termed the \emph{cosmic triad}, first proposed in \cite{Armendariz-Picon:2004say}. This scenario allows for isotropic solutions, while also accommodating situations with a preferred direction, as explored in Refs.~\cite{Orjuela-Quintana:2020klr, Motoa-Manzano:2020mwe}, considering a Bianchi I spacetime as the background. A simpler scenario involves a $U(1)$ gauge symmetry, where even general scalar-vector-tensor analyses have been conducted, for example in Refs.~\cite{Kase:2018nwt, Cardona:2022lcz}. Motivated by UV considerations, studies have delved into specific constructions, including the work of Thorsrud et al. in Ref.~\cite{Thorsrud:2012mu}, who investigated the cosmological implications of a scalar field with an exponentially field-dependent gauge kinetic function inspired by dilaton field couplings. A detailed analysis of this model was subsequently undertaken in Ref.~\cite{Orjuela-Quintana:2021zoe}, exploring alternative kinetic term structures for the scalar field.

The primary objective of our current investigation is to explore the simplest yet least constrained model inspired by string theory compactifications. Specifically, we consider two scalar fields associated with a complex modulus and a $U(1)$ vector field, where the gauge kinetic function is dependent on the modulus. The analysis leads to study a dynamical system dependent on three parameters, resulting in four fixed points.  One of these fixed points involves only one scalar field, reproducing models initially studied in this type of scenario \cite{ratraCosmologicalConsequencesRolling1988} but which is in tension with the \emph{swampland} constraints on the steepness of the potential. Two other fixed points are inherited from previous models \cite{Cicoli:2020cfj,Orjuela-Quintana:2021zoe}, which involve only a pair of fields in the model. Additionally, a completely novel fixed point is discovered in which all three fields play a role. We pay special attention to the cases where the vector field has a non-trivial role. We find that it is possible to describe the observed cosmology, albeit in a parameter space challenging to justify from superstring compactifications.

The paper has the following structure. The second section provides an overview of the construction of the scalar sector from a four-dimensional ${\cal N}=1$ Supergravity (SUGRA) theory, resulting in a generic superstring compactification. These results are then utilised to re-examine the dynamics of a cosmological system involving the modulus and its superpartner. Moving forward, the third section introduces vector fields within a SUGRA theory, proposing a contribution to the Lagrangian in the case of a $U(1)$ symmetry inspired by Heterotic and type II D-brane constructions. This model is subjected to detailed scrutiny in subsequent analyses. Then, we present our concluding remarks. The appendix shows some details about the construction of 4D effective actions from string compactifications.

\section{String inspired scalar dynamics}
\label{Sec: String inspired scalar dynamics}

\subsection{Scalar Multifield model from string compatifications}
Much in the spirit of inflationary models, our main concern will be with the scalar sector whose dynamics are expected to lead to a dynamical equation of state that hopefully presents a late-time accelerated expansion epoch. The dynamics of the scalar sector are mainly encoded in the scalar potential, which in supergravity can be split into two components: the $F$-term and $D$-term contributions. The former has the following general structure \cite{freedmanSupergravity2012}, with $M_\text{P}$ the reduced Planck mass,
\be 
V_F=e^{K/M_\text{P}^2}\left(K^{I\bar J}(D_I W)(\bar D_{\bar J} \bar W)-\frac{3}{M_\text{P}^2} W\bar W\right)\,,
\ee
where $K=K(\bar{\Phi}^{\bar I},\Phi^I)$ is the Kähler potential and $W=W(\Phi^I)$ is the superpotential. Here, $I$ and $J$ range over the complete set of chiral superfields, and the bar symbol representing the conjugate quantities; therefore, the superpotential is a holomorphic function for the chiral superfields, while the Kähler potential is a real one. Additionally, using the shorthand notation $\partial_If\equiv\frac{\partial f}{\partial \Phi^I}$, we define $K_{I\bar J}\equiv \frac{\partial^2 K}{\partial \Phi^I\partial \bar \Phi^{\bar J}}$, and its inverse $K^{I\bar K}K_{J\bar K}=\delta^I_J$. The covariant derivative is defined as
$$
D_IW= W_I+\frac{1}{M_\text{P}}K_IW\,.
$$
On the other hand, the D-term contribution has to do with the gauge symmetries,
\be 
V_D=\frac12 Re(f^{AB})D_A D_B\,,
\ee
with $f^{AB}$ the inverse of $f_{AB}$, the gauge kinetic function (more on it below), while
\be 
D_A=i X^{I}_{A}\partial_I K\,,
\ee
where $X^{I}_{A}$ are the, in general, field-dependent Killing vectors associated with the gauge symmetry $A$. In other words, a transformation associated with real parameters $\lambda^A$ induces one on the chiral superfields given by $\delta \phi^i = \lambda^AX_A^i$, such that $D_A$ is a gauge-invariant combination of charged fields.

While string compactifications exhibit a diverse array of scalar fields, the most distinctive aspect of four-dimensional effective supersymmetric theories is the presence of moduli fields, characterizing flat directions within the field space. This peculiarity motivates their incorporation into string-inspired model constructions for slow-roll inflationary and quintessence scenarios (for a review and specific instances, see \cite{Cicoli:2023opf} and references therein.) The underlying rationale for this behaviour lies in the moduli fields serving as superpartners to compactified $p$-forms potentials with gauge freedom, represented by $C_p\to C_p+\text{d}A_{p-1}$, which manifests as shift symmetries. Denoting the scalar component of our chiral modulus superfield as $\Psi=s+\rmi\sigma$ with $s$ and $\sigma$ being real, the axionic symmetry $\sigma\to \sigma+c$ is reflected in the Kähler potential as a dependency in the form
$$
K(\bar{\Phi}^{\bar I},\Phi^I)=K(\bar \Psi+\Psi,\bar{\Phi}^{\bar M},\Phi^M)=K(s,\bar{\Phi}^{\bar M},\Phi^M)\,,
$$
with $M$ ranging over all chiral fields but the modulus, while the superpotential should be modulus-independent, i.e.,
$$
W(\Phi^I)=W_0(\Phi^M)\,.
$$
In light of these considerations, the $F$-term scalar potential adopts the following expression:
\be 
V_F=e^{K(s,\bar{\Phi}^{\bar M},\Phi^M)/M_\text{P}^2}\left[ K^{M\bar N}(D_M W_0)(\bar D_{\bar N} \bar W_0)+ \frac{1}{M_\text{P}^2}\frac{|K_\Psi|^2}{K_{\Psi \bar \Psi}}|W_0|^2-\frac{3}{M_\text{P}^2}|W_0|^2\right]\,.
\ee
We further simplify by considering that the structure of the Kähler potential is of the form
\be \label{eq:factorisedK}
K(s,\bar{\Phi}^{\bar M},\Phi^M)=K(s)+{\cal K}(\bar{\Phi}^{\bar M},\Phi^M)\,.
\ee
Consequently, the scalar potential takes the following general form:
\be 
V_F=e^{K(s)/M_\text{P}^2}\left({\cal F}+{\cal G}\frac{|K_\Psi|^2}{K_{\Psi \bar \Psi}}\right)
\ee
with modulus-independent functions
$$
{\cal F}=e^{{\cal K}(\bar{\Phi}^{\bar M},\Phi^M)/M_\text{P}^2}\left(K^{M\bar N}(D_M W_0)(\bar D_{\bar N} \bar W_0)-\frac{3}{M_\text{P}^2}|W_0|^2\right)\,,
$$
and
$$
{\cal G}=e^{{\cal K}(\bar{\Phi}^{\bar M},\Phi^M)/M_P^2}\frac{|W_0|^2}{M_\text{P}^2}\,.
$$
A typical form for the modulus Kähler potential is given by a logarithmic dependency
\be 
K(s)=-p\ln\left(\frac{2s}{M_\text{P}^2}\right)M_\text{P}^2\,,
\ee
with $p$ a real constant, such that
$$
K_\Psi=-\frac{p}{2s}M_\text{P}^2\,,\quad K_{\Psi\bar\Psi}=\frac{p}{(2s)^2}M_\text{P}^2\,,
$$
and
\be 
V_F=\frac{M_\text{P}}{(2s)^p}\left({\cal F}+p{\cal G}\right)\equiv\frac{1}{s^p}{\cal V}_{0,F}\,,
\ee
with ${\cal V}_{0,F}$ a $s$-independent function. Given that ${\cal V}_{0,F}$ depends on all other fields, we examine the scenario wherein all fields except $s$ have previously been determined and stabilised through dynamics investigated over the last two decades, as illustrated for instance in  \cite{font1990supersymmetry,ferrara1990duality,Kachru:2003aw,Balasubramanian:2005zx,DeWolfe:2005uu,Camara:2005dc,Gallego:2017dvd} (for a recent review and a more complete set of references see \cite{McAllister:2023vgy}). In this context, we treat ${\cal V}_{0,F}$ as a non-vanishing constant.

Now, let us explore the $D$-term contribution. For it we restrict to the simplest case of a single $U(1)$ sector beyond the Standard Model symmetries and examine two cases. In the first case, we assume that the Killing vectors are independent of moduli and $X^\Psi=0$, indicating that $\Psi$ is neutral, resulting in
\be 
V_D=\frac{\frac12 D^2}{Re(f(\Psi))}\,,
\ee
where the numerator is $s$-independent and its value is fixed by the stabilization of all other fields. Here the dependency on $s$ will appear in the gauge kinetic function, as discussed below.

A second possibility is one of a pseudo-anomalous $U(1)$ sector, for which the field appearing as the gauge kinetic function develops a non-linear transformation with a constant Killing vector $X^\Psi=i\delta$ \cite{dineFayetIliopoulosTermsString1987,dineTermsTermsString1987,atickStringCalculationFayetiliopoulos1987}. This induces a field-dependent Fayet-Ilioupoulus term of the form
\be 
D=iX^NK_N-\delta K_\Psi=iX^NK_N-\frac{p\delta }{2s}M_\text{P}^2\,,
\ee
using the logarithmic dependency introduced above. Regarding the case on which the component $iX^NK_N=0$ once the rest of the fields are fixed\footnote{As discussed in \cite{gallegoEffectiveDescriptionLandscape2009}, the presented scenario is considered unnatural. Nonetheless, in this context, we adhere to and illustrate this particular possibility.} and the $D$-term scalar potential reduces to
\be 
V_D=\frac{(p\delta)^2 }{4s^2 Re(f(\Psi))}M_\text{P}^4\,.
\ee
In order to be explicit we use a linear dependency for the gauge kinetic function, namely $f(\Psi)=\Psi/M_P$, leading to the following general structure for the D-term scalar potential
\be 
V_D=\frac{{\cal V}_{0,D}}{s^n}M_\text{P}^{4+n}\,,
\ee
with $n=1$ and $3$ respectively in the two cases and ${\cal V}_{0,D}=\frac{1}{2M_\text{P}^4} D^2$ in the former case and ${\cal V}_{0,D}=\frac{(p\delta)^2}{4M_\text{P}^2}$ in the pseudo-anomalous one.

The kinetic part is dictated solely by the Kähler potential and is given by
\be 
{\cal L}_{kin}=-K_{I\bar J}\partial_\mu\Phi^I\partial^\mu\bar\Phi^{\bar J}\,,
\ee
thus $K_{I\bar J}$ is the metric of the field space so the consideration done in \eqref{eq:factorisedK} means that the scalar manifold in the moduli sector factorises. Using the considerations above we get the following explicit kinetic term for the scalar fields
\be 
{\cal  L}_{\text{kin}}=-\frac{pM_\text{P}^2}{(2s)^2}\partial_\mu s\partial^\mu s-\frac{pM_\text{P}^2}{(2s)^2}\partial_\mu \sigma\partial^\mu \sigma\,.
\ee
Or, defining a canonical normalised field $\phi_1=\sqrt{\frac{p}{2}}\ln(s/M_\text{P})M_\text{P}$ and renaming $\phi_2=\sigma$,
\be \label{eq:sckinstring}
{\cal  L}_{\text{kin}}=-\frac12\partial_\mu \phi_1\partial^\mu \phi_1-\frac{p}{4}e^{-2\sqrt{2/p}\phi_1/M_\text{P}}\partial_\mu \phi_2\partial^\mu \phi_2\,.
\ee
In terms of these fields the scalar potential takes the form
\be\label{eq:scpotstring}
V={\cal V}_{0,F}e^{-\sqrt{2p}\phi_1/M_\text{P}}+{\cal V}_{0,D}e^{-n\sqrt{2/p}\phi_1/M_\text{P}}\,.
\ee
This result, motivates the study of a general two-field model of the form
\be \label{eq:lagrangianscalar alone}
{\cal L}= -\frac12 \partial_\mu\phi_1\partial^\mu\phi_1-\frac12f^2(\phi_1) \partial_\mu\phi_2\partial^\mu\phi_2-V(\phi_1)\,,
\ee
with exponential functions, as was done in \cite{Cicoli:2020cfj}. In the following section, we will study the cosmological dynamics encoded in this model.

\subsection{Cosmological setup}

In a flat, homogeneous, and isotropic Friedman-Lema\^itre-Robertson-Walker (FLRW) spacetime, described by the metric
\begin{equation}
\text{d} s^2 = - \text{d} t^2 + a^2(t) \delta_{ij} \text{d} x^i \text{d} x^j\,,
\end{equation}
where $a(t)$ is the cosmic time-dependent scale factor, the Lagrangian \eqref{eq:lagrangianscalar alone} regarding homogeneous fields, yields the following equations of motion:
\begin{equation}
\label{Eq: Scalar fields eoms}
\ddot{\phi}_1 + 3 H \dot{\phi}_1 + V_{\phi_1} - f f_{\phi_1} \dot{\phi}_2^2 = 0, \qquad 
\ddot{\phi}_2 + 3 H \dot{\phi} _2 + 2 \frac{f_{\phi_1}}{f} \dot{\phi}_1 \dot{\phi}_2  = 0\,,
\end{equation}
where the dot symbol denotes time derivatives, $H \equiv \dot{a}/a$, and the subscript $\phi_1$ denotes derivatives with respect to this field. To investigate the resulting cosmology, these equations must be coupled with the Einstein equation of motion $M_\text{P}^2 G_{\mu\nu} = T_{\mu\nu}$, where $G_{\mu\nu}$ is the Einstein tensor, and $T_{\mu\nu}$ is the energy-momentum tensor. The contribution of these scalar fields to the energy-momentum tensor is expressed as
\begin{align}
T^{(\phi)}_{\mu\nu} \equiv - \frac{2}{\sqrt{-g}} \frac{\delta (\sqrt{-g}\mathcal{L})}{\delta g^{\mu\nu}}= &-\nabla_\mu \phi_1 \nabla_\nu \phi_1 - f^2(\phi_1) \nabla_\mu \phi_2 \nabla_\nu \phi_2 \\
&- g_{\mu\nu} \left( - \frac{1}{2} \nabla_\alpha \phi_1 \nabla^\alpha \phi_1 - \frac{1}{2} f^2(\phi_1) \nabla_\alpha \phi_2 \nabla^\alpha \phi_2 - V(\phi_1) \right)\,. \nonumber
\end{align}
For homogeneous fields, the density and pressure are specifically given by
\begin{equation}
\rho_\phi \equiv T_{0 0}^{(\phi)} = \frac{1}{2} \dot{\phi}_1^2 + \frac{1}{2} f^2(\phi_1) \dot{\phi}_2^2 + V, \qquad p_\phi \equiv \frac{1}{3} \text{tr}\, T_{i j}^{(\phi)} = \frac{1}{2} \dot{\phi}_1^2 + \frac{1}{2} f^2(\phi_1) \dot{\phi}_2^2 - V\,.
\end{equation}
However, for a more realistic scenario, additional components must be considered in the universe's energy budget. These contributions are represented by two perfect fluids, each following radiation and matter equations of state. Consequently, the complete energy-momentum tensor is formulated as
\begin{equation}
\label{Eq: Total EMT}
T_{\mu\nu}  = T_{\mu\nu}^{(\phi)} + T_{\mu\nu}^{(m)} + T_{\mu\nu}^{(r)}\,,
\end{equation}
and the dynamics for these two sectors adhere to the conservation of their energy-momentum tensors, i.e., $\nabla^\mu T_{\mu\nu}^{(r)} = 0$ and $\nabla^\mu T_{\mu\nu}^{(m)} = 0$, expressed explicitly as
\be
\label{Eq: continuity eqs}
\dot \rho_r + 4H \rho_r = 0\,, \qquad \dot \rho_m + 3H \rho_m = 0\,.
\ee
In light of these considerations, the Friedman equations from the Einstein field equations take the form
\begin{equation}
\label{eq: Friedman FLRW}
3 M_\text{P}^2 H^2 = \rho_\phi + \rho_m + \rho_r\,, \qquad -2 M_\text{P}^2 \dot H = \rho_\phi + p_\phi + \rho_m + \frac{4}{3} \rho_r\,.
\end{equation}

As proposed in the preceding section and elucidated in appendix \ref{app:4Dstringmodels}, superstring theory likely necessitates the coupling function and potential to conform to the expressions given in \eqref{eq:sckinstring} and \eqref{eq:scpotstring},\footnote{In the following we consider that only one of the contributions to the scalar potential \eqref{eq:scpotstring} is present. More precisely we take the D-term contribution to vanish, being at least as constrained as the F-term counterpart, as will be shown below.} explicitly represented as
\begin{equation}
\label{Eq: Coupling and Potential}
f(\phi_1) \equiv f_0 e^{-\nu \phi_1 / M_\text{P}}, \qquad V(\phi_1) \equiv V_0 e^{-\lambda \phi_1 / M_\text{P}}\,,
\end{equation}
where $f_0$ and $V_0$ are constants, and the parameters $\lambda=\sqrt{2p}$ and $\nu=\sqrt{2/p}$ are determined by the integer $p$.\footnote{In the case of the D-term scalar potential $\nu=n\sqrt{2/p}$ and a second integer is to be regarded.} Nevertheless, in the subsequent analysis, we treat $\lambda$ and $\nu$ as arbitrary and examine the parameter space of the model yielding to viable cosmologies. Subsequently, we comment on the discrete parameter space favoured by string theory.

\subsection{Dynamical Analysis}\label{sec:DynAnScaAlone}

Unveiling the cosmological dynamics of the model requires solving Eqs.~\eqref{Eq: Scalar fields eoms}, \eqref{Eq: continuity eqs} and \eqref{eq: Friedman FLRW} given some parameters and initial conditions. However, we can extract information of the asymptotic behaviour of the model encoded in its fixed points~\cite{Wainwright2009}. To do so, we recast these equations in terms of the following dimensionless variables 
\be
x_1 \equiv \frac{\dot \phi_1}{\sqrt{6} H M_\text{P}}\,,~~x_2 \equiv \frac{f(\phi_1) \dot \phi_2}{\sqrt{6} H M_\text{P}}\,,~~y \equiv \frac{\sqrt{V}}{\sqrt{3}H M_\text{P}}\,,~~\Omega_m \equiv \frac{\rho_m}{3 M_\text{P}^2 H^2}\,,~~\Omega_r \equiv \frac{\rho_r}{3 M_\text{P}^2 H^2}\,.
\label{Eq: variables}
\ee
The first Friedman equation \eqref{eq: Friedman FLRW} reduces to the constraint equation
\be
1 = x_1^2 + x_2^2 + y^2 + \Omega_m + \Omega_r\,,
\ee
and the continuity equations in Eq.~\eqref{Eq: continuity eqs} and the equations of motion for the scalar fields in Eq.~\eqref{Eq: Scalar fields eoms} are replaced by the following closed autonomous system
\bea
x_1' &=& x_1(1 + q) + \sqrt{\frac{3}{2}} \left( \lambda y^2 - 2 \nu x_2^2 - \sqrt{6} x_1 \right)\,, \label{Eq: iso x eq}\\
x_2' &=& x_2(1 + q) + x_2 \left(\sqrt{6}\nu x_1 - 3\right)\,, \\
y' &=& y \left(1 + q - \frac{\sqrt{6}}{2} \lambda x_1 \right)\,, \\
\Omega_r' &=& 2 \left(q - 1\right) \Omega_r\,, \label{Eq: iso r eq}
\eea
where a prime denotes a derivative with respect to the number of $e$-folds, which is related to the cosmic time by $\text{d}N \equiv H \text{d}t$. The deceleration parameter, $q \equiv - a \ddot{a}/\dot{a}^2$, is written in terms of these variables as
\bea\label{eq:qparameter}
1 + q = -\frac{\dot{H}}{H^2} = \frac{1}{2} \left(3 x_1^2 + 3 x_2^2 - 3 y^2 + \Omega_r + 3\right)\,.
\eea
The fixed points of this system are computed by solving the algebraic system resulting from $x_1' = 0$, $x_2' = 0$, $y' = 0$, and $\Omega_r' = 0$. We characterise the physical meaning of these fixed points by computing the effective equation of state and the equation of state of dark energy
\begin{equation}
\label{Eq: Iso weff}
w_\text{eff} \equiv \frac{p_\text{tot}}{\rho_\text{tot}} = -1 - \frac{2}{3} \frac{\dot{H}}{H^2}\,, \qquad w_\text{DE} \equiv \frac{p_{\phi}}{\rho_{\phi}} = -1 + \frac{2\left(x_1^2 + x_2^2 \right)}{x_1^2 + x_2^2 + y^2}\,,
\end{equation}
where $\rho_\text{tot} \equiv \rho_\phi + \rho_m + \rho_r$ and $p_\text{tot} = p_\phi + \rho_r/3$.
Here, we focus on accelerated solutions driven by dark energy, i.e., fixed points where $w_\text{eff} < -1/3$ and $w_\text{DE} < - 1/3$. Additionally, notice that the autonomous system exhibits the following symmetries:
\begin{equation}
 \{x_1, \lambda, \nu \} \rightarrow \{-x_1,  -\lambda, -\nu \}\,, \quad x_2 \rightarrow -x_2\,, \quad y \rightarrow -y\,.
\end{equation}
Therefore, without loss of generality we restrict ourselves to solutions with ${x_2, y} \geq 0$ and consider non-negative values of the parameters $\lambda$ and $\nu$ while leaving free $x_1$. Under these conditions we obtained just two possible accelerated solutions:
\begin{align}
\mathcal{G}&: x_1 = \frac{\lambda}{\sqrt{6}}\,, \quad x_2 = 0\,, \quad y = \sqrt{1 - \frac{\lambda^2}{6}}\,, \quad \Omega_r = 0\,, \quad w_\text{DE} = -1 + \frac{\lambda^2}{3}\,, \label{Eq: FP iso 1}\\
\mathcal{NG}&: \label{Eq: FP iso 2} \\
x_1 &= \frac{\sqrt{6}}{\lambda + 2\nu}\,, \ x_2 = \frac{\sqrt{\lambda^2 + 2\nu\lambda - 6}}{\lambda + 2\nu}\,, \ y = \sqrt{\frac{2\nu}{\lambda + 2\nu}}, \ \Omega_r = 0\,, \ w_\text{DE} = -1 + \frac{2\lambda}{\lambda + \nu}\,. \nonumber
\end{align}
\begin{figure}[t!]
\includegraphics[width = 0.48\textwidth]{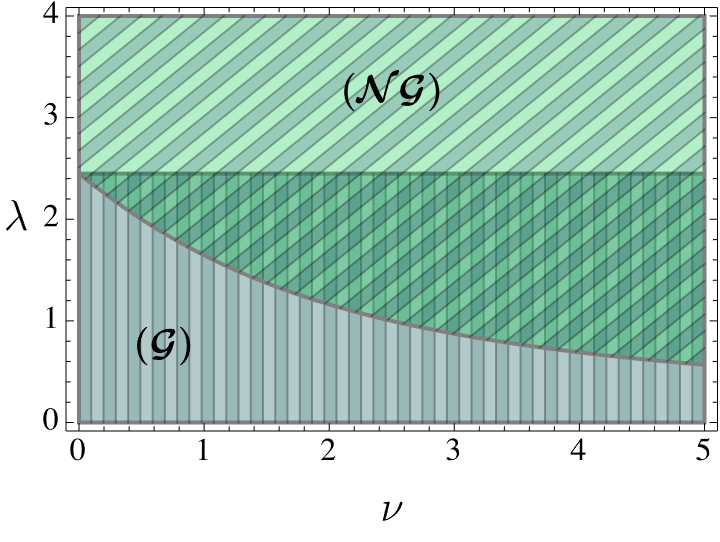}
\hfill
\includegraphics[width = 0.48\textwidth]{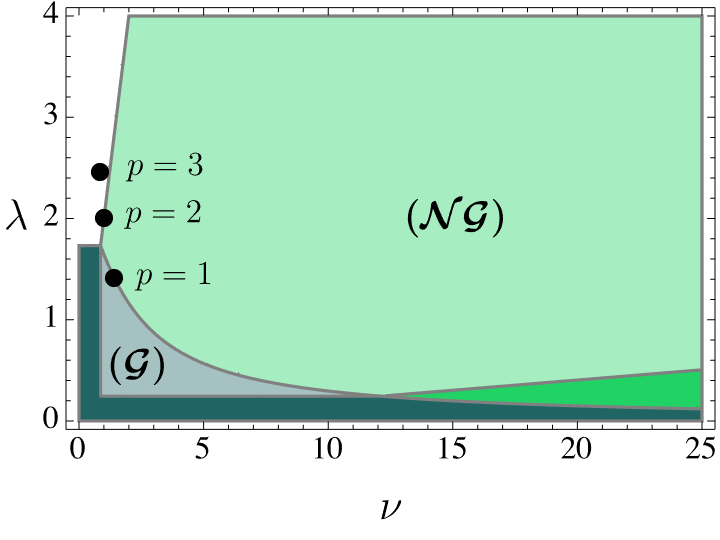}
\caption{(Left) Existence (real valued variables) regions of ($\mathcal{G}$) and ($\mathcal{NG}$) in the parameter space $\{\nu, \lambda\}$. (Right) Attraction regions of these points. The darker regions correspond to the regions where $w_\text{DE} \sim -1$, which in general require $\nu \gg \lambda$ for both points, although ($\mathcal{G}$) can be a suitable accelerated attractor for any $\nu$ given that $\lambda \ll 1$ as in the usual quintessence scenario. We can see that the values of $\lambda$ and $\nu$ predicted by string theory happen to fall outside the region associated with stable and viable accelerated attractors.}
\label{Fig: Regions QS Model}
\end{figure}
These points were called ``geodesic'' ($\mathcal{G}$) and ``non-geodesic'' ($\mathcal{NG}$) solutions in Ref.~\cite{Cicoli:2020cfj}, since they are related to the ``curvature'' in the field space of the scalar fields.  

In the left panel of figure~\ref{Fig: Regions QS Model}, we show the existence region of the points ($\mathcal{G}$) and ($\mathcal{NG}$), defined as the parameter space regions $\{\nu, \lambda\}$ where the variables in the fixed points [Eqs.~\eqref{Eq: FP iso 1} and \eqref{Eq: FP iso 2}] attain real values. On the other hand, the stability of a fixed point can be known by analysing the behaviour of small perturbations of the autonomous system around the given fixed point. At the linear level, this amounts to evaluate the real part of the eigenvalues of the Jacobian matrix of the autonomous system in the corresponding fixed points. If all eigenvalues are negative, it is said that the point is an attractor. If there is a mix of negative and positive eigenvalues, then we have a saddle fixed point, and if all eigenvalues are positive, this point is a source~\cite{Wainwright2009}. In the right panel of figure~\ref{Fig: Regions QS Model}, we show the parameter space where these points emerge as attractors of the system, i.e., their corresponding regions of attraction. We particularly emphasize the subregions associated with an accelerated epoch, meeting the condition $-1 \leq w_\text{eff} \leq -0.98$. Recalling that $\lambda = \sqrt{2 p}$ and $\nu = \sqrt{2/p}$ are demanded by string theory, we depict the points for some values of $p$, observing that none of them describes a scenario of late-time accelerated expansion in agreement with current observations. This clearly indicates that the current state of the universe is challenging to explain within the framework of this model, as derived from string compactifications. To grasp this, observe that as $p$ increases, the steepness of the potential for $\phi_1$ also intensifies, necessitating a more pronounced interplay with $\phi_2$, given by a larger $\nu$ parameter, as depicted in the figure. However, $\nu$ decreases with increasing $p$, creating a tension between these two requirements. This possibility was identified in Ref.~\cite{Cicoli:2020cfj} as a regime of ``large curvature'' in the scalar fields space. However, as illustrated in Figure \ref{Fig: Regions QS Model}, the parameter values required by string theory lie outside the regions where the model exhibits viable cosmological attractors.

\section{Scalar multifield coupled to a vector field}
\label{Sec: Scalar multifield coupled to a vector field}

While currently challenging to justify the existence of accelerated solutions in the multifield model that align with observations, alternative mechanisms may potentially broaden the model's parameter space. For example, in Refs. \cite{Alvarez:2019ues, Orjuela-Quintana:2020klr, BeltranAlmeida:2019fou, Orjuela-Quintana:2021zoe, Akrami:2020zfz}, it was demonstrated that steep potentials, such as Higgs-like or exponential potentials resembling that in Eq.~\eqref{Eq: Coupling and Potential} with $\lambda^2 > 2$, can yield stable accelerated solutions when supported by the presence of other fields, including non-abelian $SU(2)$ vector fields, abelian $U(1)$ vector fields, or 2-form fields. In this section, we explore how this possibility holds true for the scalar multifield model.

\subsection{Cosmological setup}

Here, inspired by Ref.~\cite{Orjuela-Quintana:2021zoe} where it was shown that a vector field coupled to a scalar field yields anisotropic accelerated attractors even for a steep exponential potential, we explore a scenario involving a coupling between the scalar field responsible for the accelerated expansion, denoted as $\phi_1$, and a vector field represented by $A_\mu$. The Lagrangian of this extended model reads 
\begin{equation}
\label{Eq: lagrangian with A}
\mathcal{L} = -\frac{1}{2} \partial_\mu \phi_1 \partial^\mu \phi_1 - \frac{1}{2} f^2(\phi_1) \partial_\mu \phi_2 \partial^\mu \phi_2 - V(\phi_1) - \frac{1}{4} h^2(\phi_1) F_{\mu\nu} F^{\mu\nu}\,,
\end{equation}
where $F_{\mu\nu} \equiv \nabla_\mu A_\nu - \nabla_\nu A_\mu$ is the strength tensor associated to the vector field, and $h^2(\phi_1)$ is the field dependent coupling to the gauge kinetic function. First of all, the energy-momentum tensor in Eq.~\eqref{Eq: Total EMT} is extended by a contribution from the gauge sector which reads
\be
T^{(A)}_{\mu\nu} = h^2(\phi_1) \left[ F_{\mu\alpha} F_{\ \nu}^{\alpha} - \frac{1}{4} g_{\mu\nu} F_{\alpha\beta} F^{\alpha\beta} \right]\,.
\ee
In a FLRW spacetime, all the components of $F_{\mu\nu}$ are zero since only a time-dependent time-like vector field keeps the isotropy of the background~\cite{Koivisto:2008xf}. Nonetheless, there is a recent increase in the amount of evidence suggesting that the Universe could not be as isotropic as thought \cite{Secrest:2020has, Secrest:2022uvx, Dam:2022wwh, Perivolaropoulos:2021jda, Jones:2023ncn}. Therefore, as in Ref.~\cite{Orjuela-Quintana:2021zoe}, we adopt an anisotropic Bianchi-I metric with a residual symmetry in the $y$-$z$ plane\footnote{This selection, chosen for its simplicity, is not lacking in generality. In fact, as demonstrated in Ref.~\cite{Karciauskas:2016pxn}, it was shown that the most general configuration for the space-like vector field and the Bianchi-I metric can be effectively reduced to an axially symmetrical system. The axial symmetry is determined by the perpendicular component of the vector field, while its parallel component defines the residual symmetry plane.} and a spatial-like vector field with only one component aligned with the $x$ axis. Explicitly 
\begin{equation}
\label{Eq: Bianchi metric}
\text{d} s^2 = -\text{d} t^2 + a^2(t) \left[ e^{-4\sigma} \text{d}x^2 + e^{2\sigma} ( \text{d} y^2 + \text{d} z^2 ) \right]\,, \qquad A_\mu = (0, A(t), 0, 0)\,,   
\end{equation}
where $a(t)$ in this case represents an overall scale factor, and the anisotropy is encoded in the geometric shear $\sigma(t)$. Using this metric and this vector field profile, we distinguish the vector field contributions to the density and pressure
\begin{equation}
\label{eq:vectorfieldenden}
\rho_A = \frac{1}{2} h^2(\phi_1) \frac{\dot{A}^2 e^{4\sigma}}{a^2}\,, \qquad p_A = \frac{1}{3} \rho_A\,.
\end{equation}
From the Einstein equations, besides the Friedman equations which will take into account the contribution of the vector field to the stress-energy tensor, we have the corresponding equation for the evolution of the geometric shear\footnote{An equation of motion for the shear can be derived from the difference between the ``$11$'' and ``$22$'' components of the Einstein equations, i.e., $M_P^2(G_{\ 2}^2-G_{\ 1}^1)=T_{\ 2}^2-T_{\ 1}^1$.}
\bea
3M_\text{P}^2 H^2 &=& \rho_A + \rho_\phi + \rho_m + \rho_r + 3M_\text{P}^2 \dot{\sigma}^2\,, \label{Eq: Ani 1º Friedman Eq}\\
-2 M_\text{P}^2 \dot{H} &=& \frac{4}{3} \rho_A + \dot{\phi}_1^2 + f^2 \dot{\phi}_2^2 + \rho_m + \frac{4}{3} \rho_r + 6 M_\text{P}^2 \dot{\sigma}^2\,, \\
\ddot \sigma &=& - 3H\dot \sigma + \frac{2}{3M_\text{P}^2}\rho_A\,. \label{eq:sheareom}
\eea
In the last equation, we note that in the case $\rho_A = 0$, the shear decays as $\sigma \propto a^{-3}$, i.e., the anisotropy of the Universe is erased by the expansion. This resembles the well-known no-cosmic hair theorem valid for $\Lambda$CDM~\cite{Wald:1983ky}. Instead, in our case, the density of the vector field sources the shear preventing this fast decay. However, for a phenomenologically viable scenario, the predicted values must conform to the current observational constraints~\cite{Campanelli:2010zx, Amirhashchi:2018nxl}.

Since the vector field is not coupled to $\phi_2$, the equation of motion of this scalar field is not modified [see Eq.~\eqref{Eq: Scalar fields eoms}]. In turn, the equation of motion for $\phi_1$ and $A_\mu$ are given by
\bea
\ddot{\phi}_1 + 3 H \dot{\phi}_1 + V_{\phi_1} - f f_{\phi_1} \dot{\phi}_2^2 - 2\frac{h_{\phi_1}}{h} \rho _A &=& 0\,, \\
\ddot{A} + (H + 4 \dot{\sigma}) \dot{A} + 2 \frac{h_{\phi_1}}{h} \dot{\phi}_1 \dot{A} &=& 0\,.
\eea
In the following section we perform the analysis of this enriched dynamics.

\subsection{Dynamical Analysis}

To investigate the asymptotic evolution of the fully-fledged model, we introduce two extra dimensionless variables to the set outlined in Eq.~\eqref{Eq: variables}. Furthermore, we adopt an exponential functional form for the gauge kinetic function, a choice motivated by results from string compactifications, as elucidated later on. Thus we define
\begin{equation}
z^2 \equiv \frac{\rho_A}{3 M_\text{P}^2 H^2}, \qquad \Sigma \equiv \frac{\dot{\sigma}}{H}, \qquad h(\phi_1) \equiv h_0 e^{-\mu \phi_1 / M_\text{P}}\,,
\end{equation}
where $h_0$ and $\mu$ are constants. The first Friedman equation in Eq.~\eqref{Eq: Ani 1º Friedman Eq} becomes the constraint
\be
1 = x_1^2 + x_2^2 + y^2 + z^2 + \Omega_m + \Omega_r + \Sigma^2\,.
\ee
Concerning the autonomous system in Eqs.~\eqref{Eq: iso x eq}-\eqref{Eq: iso r eq}, we observe that only the equation for the variable $x_1$ undergoes a modification in its structure. Additionally, two more equations must be incorporated to account for the introduction of new variables, namely $z$ and $\Sigma$. We have that the complete set is
\begin{align}
x_1' &= x_1(1 + q) + \sqrt{\frac{3}{2}} \left( \lambda y^2 - 2\mu z^2 - 2 \nu x_2^2 - \sqrt{6} x_1 \right)\,, \label{Eq: ani x eq} \\
x_2' &= x_2\left(1 + q + \sqrt{6}\nu x_1 - 3\right), \label{eq:x2eomV}\\
y' &= y \left(1 + q - \frac{\sqrt{6}}{2} \lambda x_1 \right)\,, \\
z' &= z\left\{(q - 1) +\sqrt{6}\mu x_1 -2 \Sigma \right\}\,, \\
\Sigma' &= \Sigma \left(q - 2\right) + 2 z^2\,, \\
\Omega_r' &= 2 \left(q - 1\right) \Omega_r\,, \label{Eq: ani r eq}
\end{align}
where we used the equation of motion for the vector field, which we have conveniently recast in terms of the density $\rho_A$ as
\begin{table}[t]
\centering
$$
\begin{array}{c|cccccc|c}\hline
\text{FP} & x_1 & x_2 &  y & z & \Sigma& \Omega_r & w_\text{eff} \\
\hline \hline
\text{(DE1)} & \frac{2\sqrt{6}\mathcal{A}}{8 + \mathcal{A}\mathcal{C}} & 0 & \frac{\sqrt{6(8 - \mathcal{A}\mathcal{B})(2 + \mu\mathcal{A})}}{8 + \mathcal{A}\mathcal{C}} &  \frac{\sqrt{3(8 - \mathcal{A}\mathcal{B})( \lambda\mathcal{A} - 4)}}{8 + \mathcal{A}\mathcal{C}} & \frac{2(\lambda\mathcal{A} - 4)}{8 + \mathcal{A}\mathcal{C}} & 0 & -1 + \frac{4 \lambda \mathcal{A}}{8 + \mathcal{A}\mathcal{C}}\\
& & & & & & & \\
\text{(DE2)} & \frac{\sqrt{6}}{\lambda +2 \nu } & \frac{\sqrt{3} \sqrt{{\cal A} ({\cal B}+4 \nu )-8}}{2 (\lambda +2 \nu )} & \frac{\sqrt{\frac{3}{2} \nu (\lambda -2 \mu +4 \nu)}}{\lambda +2 \nu } &  \frac{\sqrt{\frac{3}{2}\nu ({\cal C}-4 \nu )}}{\lambda +2 \nu } & \frac{{\cal C}-4 \nu }{2 \lambda +4 \nu } & 0 & -1 + \frac{2 \lambda }{\lambda +2 \nu } \\
\hline
\end{array}
$$
\caption{Values of the variables in the anisotropic accelerated solutions (DE1) and (DE2). We have defined ${\cal A} \equiv (\lambda +2 \mu )$, ${\cal B} \equiv \lambda-6\mu$, ${\cal C} \equiv \lambda+6\mu$ to keep the presentation simple.}
\label{tab:fpred}
\end{table}
\be
\dot \rho_A + 4\left(H + \dot \sigma\right)\rho _A + 2\frac{ \dot h_{\phi_1}}{h}\dot \phi _1\rho _A = 0\,.
\ee
The deceleration parameter $1 + q \equiv - \dot{H}/H^2$ is given by
\bea\label{Eq: ani q parameter}
1 + q = \frac{1}{2} \left(3 x_1^2 + 3 x_2^2 - 3 y^2 + \Omega_r + z^2 + 3\Sigma^2 + 3\right)\,.
\eea
In this case, the effective equation of state undergoes corrections due to the presence of shear. It can be computed as
\begin{equation}
\label{Eq: Ani weff}
w_\text{eff} = - 1 - \frac{2}{3} \frac{\dot{H}}{H^2} \frac{1}{1 - \Sigma^2} - 2 \frac{\Sigma^2}{1 - \Sigma^2}\,,
\end{equation}
Together with the scalar and vector fields, it is convenient to include the shear in the definition of the dark energy fluid, as follows:
\begin{equation}
\rho_\text{DE} \equiv \rho_\phi + \rho_A + 3 M_\text{P}^2 \dot{\sigma}^2, \qquad p_\text{DE} \equiv p_\phi + p_A + 3 M_\text{P}^2 \dot{\sigma}^2\,,
\end{equation}
such that the equation of state of dark energy is 
\begin{equation}
w_\text{DE} \equiv \frac{p_\text{DE}}{\rho_\text{DE}} = -1 + \frac{2}{3} \frac{3 (x_1^2 + x_2^2 + \Sigma^2) + 2 z^2}{x_1^2 + x_2^2 + y^2 + z^2 + \Sigma^2}\,.
\end{equation}
The autonomous system exhibits the following symmetries:
\begin{equation}
 \{x_1, \lambda, \nu, \mu \} \rightarrow \{-x_1,  -\lambda, -\nu, -\mu \}, \quad x_2 \rightarrow -x_2, \quad y \rightarrow -y, \quad z \rightarrow -z.
\end{equation}
Hence, without loss of generality, we confine our exploration to solutions with positive values for $x_2$, $y$, and $z$, along with non-negative parameters $\lambda$ and $\nu$, while leaving $x_1$ and $\mu$ unrestricted. Under these conditions, we obtained the isotropic solutions ($\mathcal{G}$) and ($\mathcal{NG}$), as discussed in section~\ref{sec:DynAnScaAlone}. Additionally, we identified two potential anisotropic accelerated solutions, named (DE1) and (DE2). The variable values at these points are presented in Table~\ref{tab:fpred}. Notably, point (DE1) aligns precisely with the anisotropic accelerated solution discovered in Ref.~\cite{Orjuela-Quintana:2021zoe} for the quintessence model. This correspondence is expected since $x_2=0$ is an automatic solution to \eqref{eq:x2eomV}, reducing the system to a case where $\phi_2$ is absent. On the other hand, point (DE2) corresponds to a clear generalisation of point ($\mathcal{NG}$), considering anisotropy.

\subsubsection{Fixed Point (DE1)}

\begin{figure}[t!]
\centering
 \includegraphics[width =0.49\textwidth]{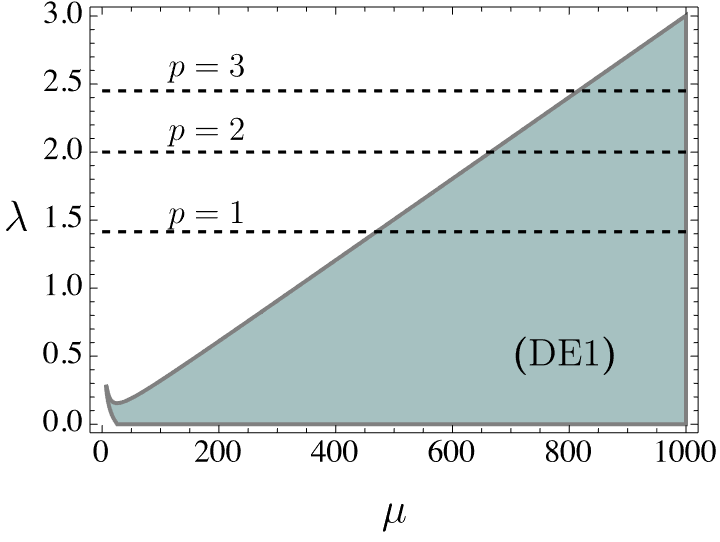}
\hfill
\includegraphics[width = 0.47\textwidth]{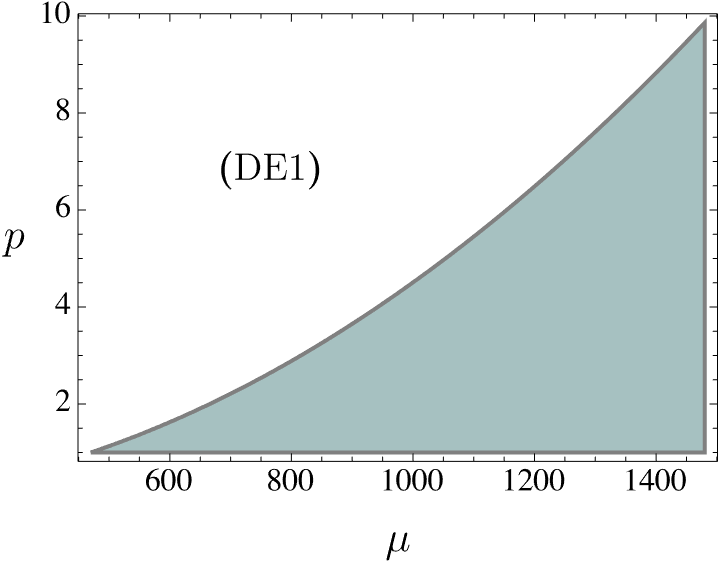}
\caption{(Left) Existence region of the point (DE1). The dashed lines represent $\lambda = \sqrt{2p}$ for some values of $p$. Viable solutions [$w_\text{DE} \sim -1$ and $\Sigma$ small as in Eq.~\eqref{Eq: cosmo viable}] within the string theory requirements ($\lambda = \sqrt{2p}$) imply $\mu \gg \lambda$. Note that the value of $\nu$ is irrelevant for the existence of this fixed point since $x_2 = 0$ in (DE1). (Right) Stability region of (DE1) considering $\mu \gg \lambda$, $\lambda = \sqrt{2p}$ and $\nu = \sqrt{2/p}$ in the approximated expresions for the eigenvalues $\eta_{1, \ldots, 6}$ in Eq.~\eqref{Eq: DE1 approximations}. We can see that this point can provide viable anisotropic accelerated solutions adjusting the values of $\mu$ to the given discrete value of $p$.
}
\label{Fig: Regions DE1}
\end{figure}
In left panel of Figure~\ref{Fig: Regions DE1}, we plot the existence region for (DE1) considering that
\begin{equation}
\label{Eq: cosmo viable}
-1 \leq w_\text{DE} < -0.98, \qquad -0.001 \leq \Sigma \leq 0.001,
\end{equation}
where the bounds for $\Sigma$ are within the constraints given in Refs.~\cite{Campanelli:2010zx, Amirhashchi:2018nxl}. We also plot three lines denoting the values of $\lambda$ for $p = 1$, $2$, and $3$. We see that in general, the existence of anisotropic accelerated solutions in agreement with observations require $\lambda \ll \mu$. Under this condition, we see that
\begin{equation}
w_\text{eff} = w_\text{DE} \approx -1 + \frac{2}{3}\frac{\lambda}{\mu}, \qquad \Sigma \approx \frac{\lambda}{3\mu} - \frac{2}{3 \mu^2}.
\end{equation}
Also, it is possible to estimate the eigenvalues of the Jacobian matrix in this point. We get
\begin{equation}
\label{Eq: DE1 approximations}
\eta_1 \approx -4 + \frac{2\lambda}{\mu}, \quad \eta_2 \approx -3 + \frac{\lambda + 2\nu}{\mu}, \quad \eta_{3, 4} \approx - 3 - \frac{7\lambda}{2\mu}, \quad \eta_{5, 6} \approx -\frac{3}{2} - \frac{7\lambda}{2\mu}.
\end{equation}
From the latter expressions, we can estimate the attraction region of (DE1), i.e., the region where $\eta_{1, \ldots, 6} < 0$. We get two branches depending on the sign of $\mu$:
\begin{equation}
\mu < 0: \ \{\lambda < -3\mu/7, \ \nu > 0\}, \qquad \mu > 0: \ \{\lambda < 2\mu, \ 2\nu < 3\mu - \lambda\}.
\end{equation}
However, when we consider the parameter space for viable cosmological solutions [Eq.~\eqref{Eq: cosmo viable}], the positive branch is dynamically selected and $\mu \gg \lambda$ implies that the cosmological interesting attraction region is delimited by
\begin{equation}
\mu > 2\nu/3.
\end{equation}


Notably, when $\lambda = \sqrt{2p}$ and $\nu = \sqrt{2/p}$, we observe in Figure~\ref{Fig: Regions DE1} that (DE1) serves as an accelerated attractor of the system, for some values of $p$ within the range expected from string theory and large values of $\mu$. For instance, for $p=1$, the viable cosmological attractor require $\mu \gtrsim 496$. This can be physically interpreted as the vector field interaction, controlled by $\mu$, playing a crucial role in inducing a slow-roll behavior in the multifield configuration evolution, despite the presence of a steep potential. 

\subsubsection{Fixed Point (DE2)}

From the expressions in Table~\ref{tab:fpred}, we see that for (DE2) to be a cosmologically viable point, i.e., $w_\text{DE} \sim -1$ and $\Sigma$ small [see Eq.~\eqref{Eq: cosmo viable}], we require $\nu \gg \lambda$. Under this condition, we get
\begin{equation}\label{eq:omegashearDE2}
w_\text{eff} \approx -1 + \frac{\lambda}{\nu}, \qquad \Sigma \approx -1 + \frac{3}{2}\frac{\mu}{\nu}+ \frac{3}{4}\frac{\lambda}{\nu} \left( \frac{\mu-\nu}{\nu}\right).
\end{equation}
Hence, the admissible solutions for any $\mu$ dismiss the values $\lambda = \sqrt{2p}$ and $\nu = \sqrt{2/p}$ stemming from string theory. Nevertheless, in the subsequent analysis, we extensively survey the parameter space of this novel solution, allowing for arbitrary values of these parameters.

Concerning (DE2), the eigenvalues of the Jacobian matrix manifest as highly intricate expressions, rendering an analytical approach challenging to implement. Consequently, we opt for a numerical assessment of the real part of the eigenvalues of the Jacobian matrix at these points, employing a methodology akin to that proposed in Ref.~\cite{Garcia-Serna:2023xfw}. 

\begin{figure}[t!]
\includegraphics[width = 0.51\textwidth]{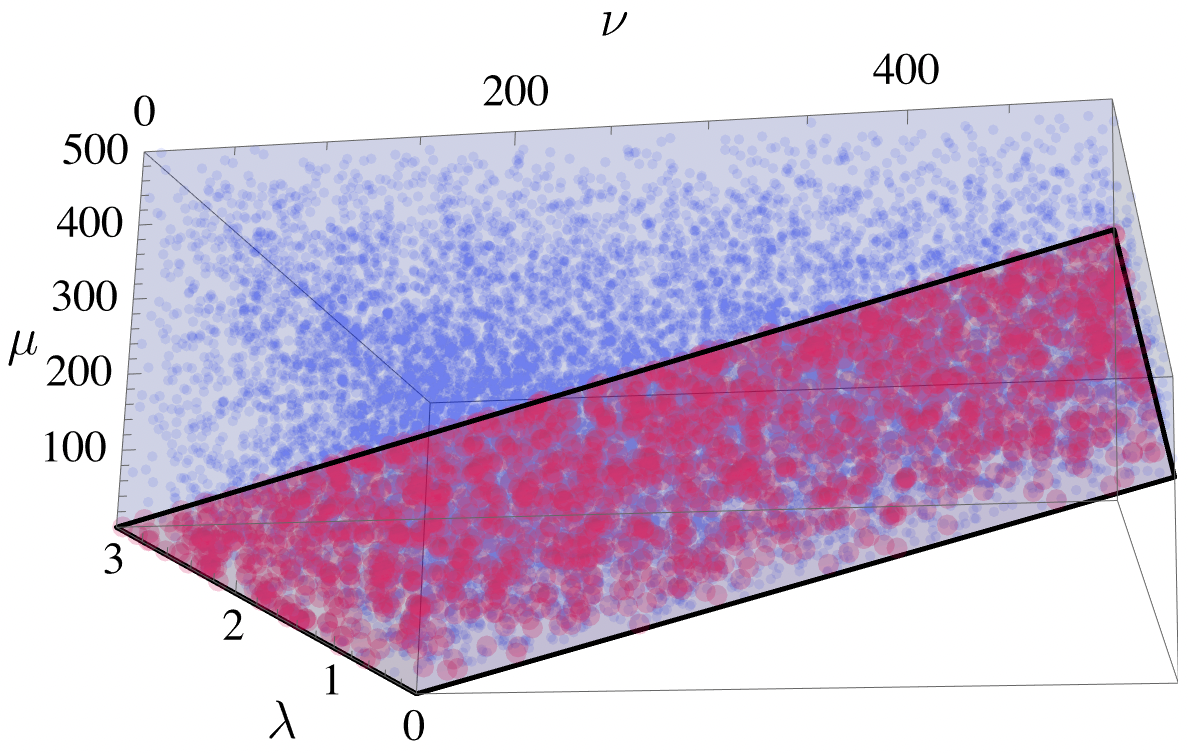}
\hfill
\includegraphics[width = 0.47\textwidth]{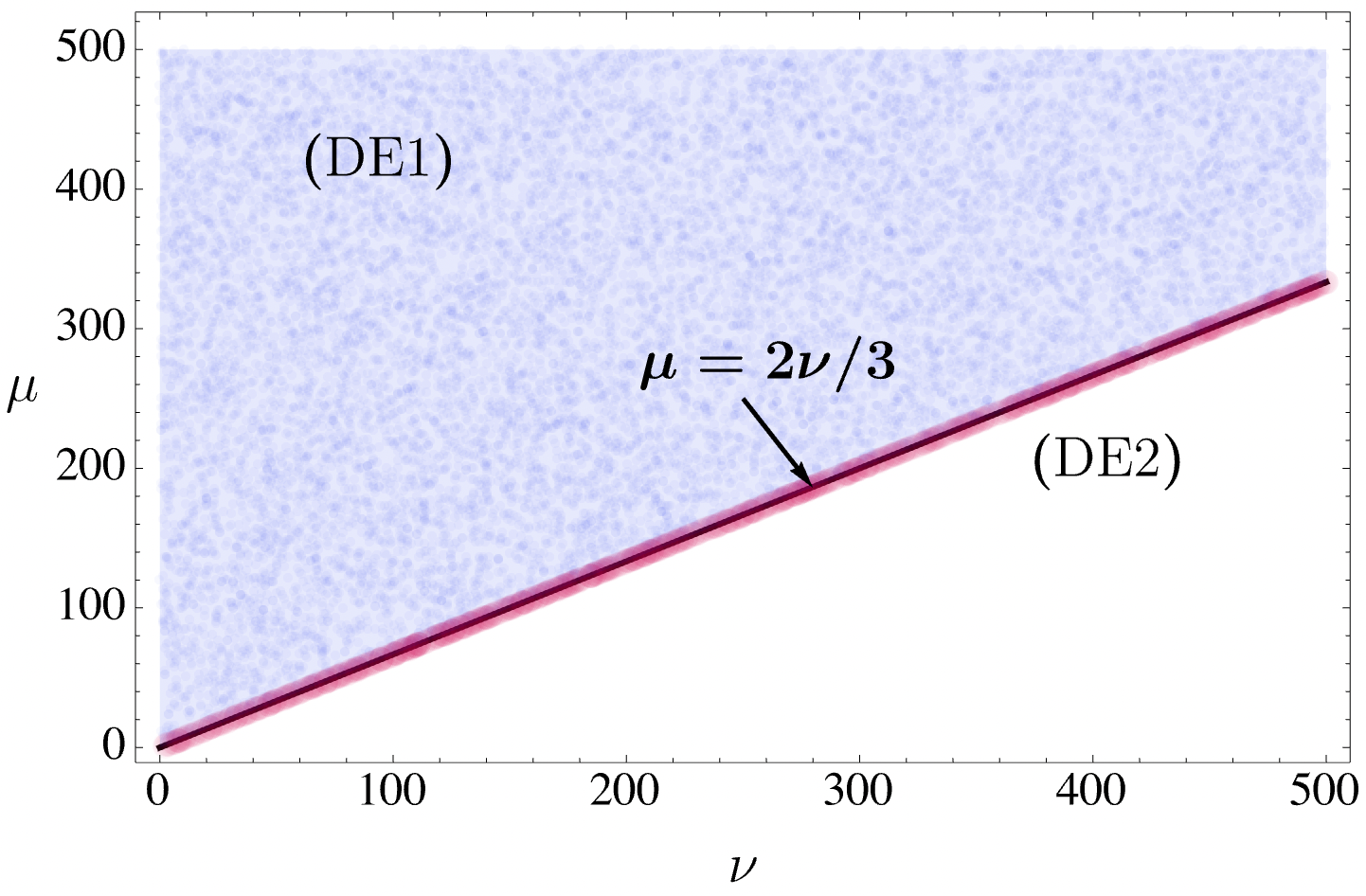}
\caption{Attraction regions of the points (DE1) (blue) and (DE2) (red), for $\lambda \in [0, 3]$, and $\{\nu, \mu\} \in [0, 500]$. The solid lines correspond to the boundaries of the $\mu = 2\nu/3$ plane, ruling the (DE2) attractor points. This is further supported by projecting onto the $\{\nu, \mu\}$ plane in the right panel.}
\label{Fig: Bifurcation Curve}
\end{figure}

\begin{figure}[t!]
\centering
\includegraphics[width = 0.58\textwidth]{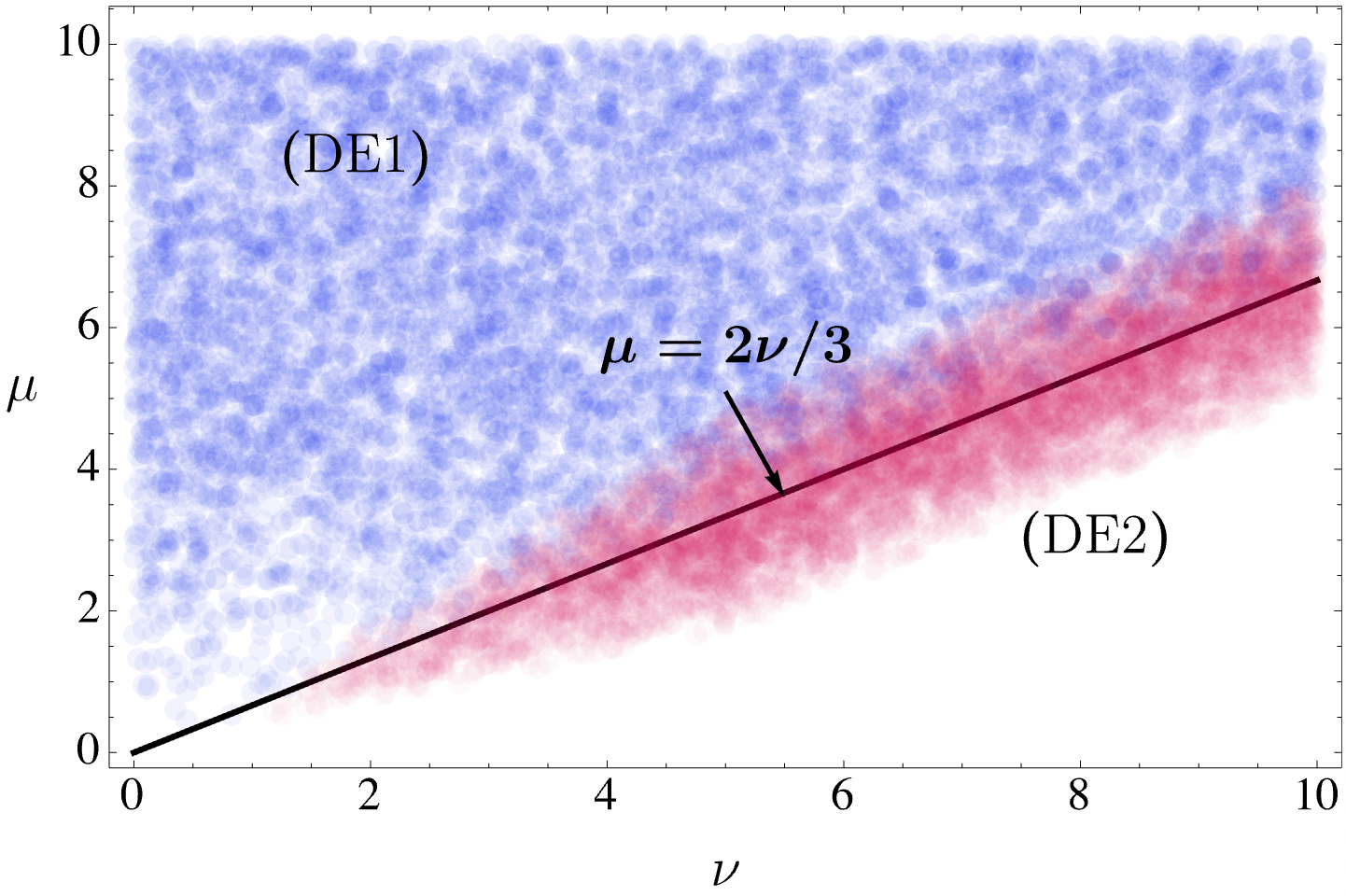}
\caption{Similar to Figure~\ref{Fig: Bifurcation Curve} but considering a smaller parameter region to perform the random sampling, namely, $\{\lambda, \nu, \mu\} \in [0, 10]$. All the (DE2) attractors are near the plane $\mu = 2\nu/3$ but not necessarily into this plane as it seems in Figure~\ref{Fig: Bifurcation Curve}. We corroborated that the overlap between blue and red points is merely due to the projection onto $\lambda$, and that a proper bifurcation manifold between (DE1) and (DE2) exists in the complete parameter space $\{\lambda, \nu, \mu\}$.}
\label{Fig: Close Bifurcation Curve}
\end{figure}

The numerical methodology can be succinctly encapsulated in the following manner. A large number of random points is sampled within the parameter space $\{ \lambda, \nu, \mu \}$, and the fixed points are assessed at these locations. Subsequently, only the parameter sets corresponding to real-valued fixed points are retained. The eigenvalues of the Jacobian matrix are then computed for these specific parameter sets, and those associated with attractor solutions are selected. In figures~\ref{Fig: Bifurcation Curve} and \ref{Fig: Close Bifurcation Curve}, we show the result of this procedure when applied to the points (DE1) and (DE2). 

In Figure~\ref{Fig: Bifurcation Curve}, we explored the parameter space by varying $\lambda$ from 0 to 3, and $\{\nu, \mu\}$ from 0 to 500. The left panel of the figure illustrates that all the blue points, corresponding to locations where all eigenvalues evaluated at (DE1) are negative, are allocated above the plane $\mu = 2\nu/3$. Conversely, the red points, indicating positions where all eigenvalues at (DE2) are negative, align with the plane described by $\mu = 2\nu/3$. This observation is further supported by analysing the projection of the stability regions onto the $\{\nu, \mu\}$ plane, as depicted in the right panel of Figure \ref{Fig: Bifurcation Curve}. 

To assess scenarios where all parameters are of the same order, we conducted a more focused sampling within the parameter space, specifically $\{\lambda, \nu, \mu\} \in [0, 10]$. The outcomes are presented in Figure~\ref{Fig: Close Bifurcation Curve}. In this case, the projection onto the $\{\nu, \mu\}$ plane reveals a mixture of attractor locations between (DE1) and (DE2). However, this should be expected since we are considering a projection while there is a mild dependency on $\lambda$. We numerically corroborate that a proper bifurcation manifold exists in the complete parameter space $\{\lambda, \nu, \mu\}$. We also notice that (DE2) attractors (red points) are situated near the plane $\mu = 2\nu/3$, but not necessarily into this plane as it seems in Figure~\ref{Fig: Bifurcation Curve}.

Intriguingly, the attractor nature of (DE2) emerges at points where the anisotropy is naturally suppressed, i.e., $\mu\approx 2\nu/3$, while, simultaneously, exhibits a late-time accelerated phase, implying $\lambda\ll \nu$, as evident from equation \eqref{eq:omegashearDE2}. However, this observation also implies that the parameter $\mu$, which governs the gauge dynamics, should be finely tuned in relation to scalar couplings controlled by $\nu$, while also resulting in a subdominant contribution of the vector field to the dark energy budget.

This final observation leads us to the conclusion that, although both (DE1) and (DE2) can portray a scenario explaining the current observations, (DE2) is much less likely to occur compared to (DE1) when the complete model is taken into account. This is evident in Figures~\ref{Fig: Bifurcation Curve} and \ref{Fig: Close Bifurcation Curve}, where the parameter space for (DE1) is significantly larger than that for (DE2).

\subsection{Cosmological Dynamics}

In this section, we numerically solve the autonomous set in Eqs.~\eqref{Eq: ani x eq}-\eqref{Eq: ani r eq} for a particular set of parameters. The chosen parameters are tailored to make (DE1) the attractor point, with a subsequent discussion on the situation for (DE2). Specifically, we opt for $\lambda = \sqrt{2p}$, $\nu = \sqrt{2/p}$, where $p = 2$, and $\mu = 10^3$. The initial conditions are defined as follows:
\begin{equation}
\label{Eq: init cond}
x_{1, i} = x_{2, i} = 10^{-15}\,, \quad y_i = 2 \times 10^{-14}\,, \quad z_i = 10^{-12}\,, \quad \Sigma_i = 0\, \quad \Omega_{r_i} = 0.99995\,.
\end{equation}
These conditions represent a radiation-dominated era, initiated at $z_{r, i} = 6.5 \times 10^{7}$, corresponding to $N = -\log{(1 + z)} = -18$. Given our interest in late-time mechanisms for generating an anisotropic expansion, we assume an initially smooth Universe, i.e., $\Sigma_i = 0$.
\begin{figure}[t!]
\includegraphics[width = 0.51\textwidth]{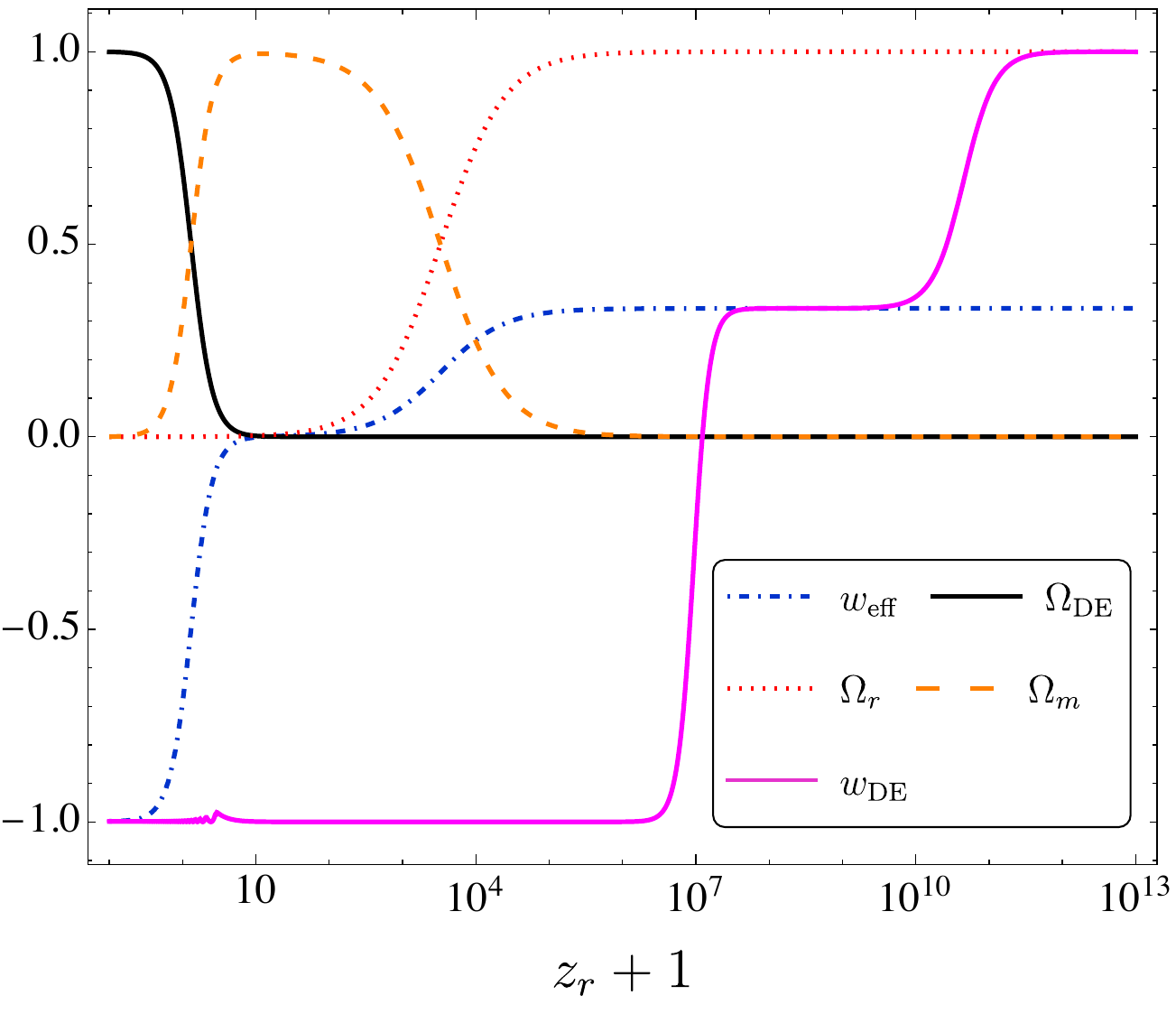}
\hfill
\includegraphics[width = 0.43\textwidth]{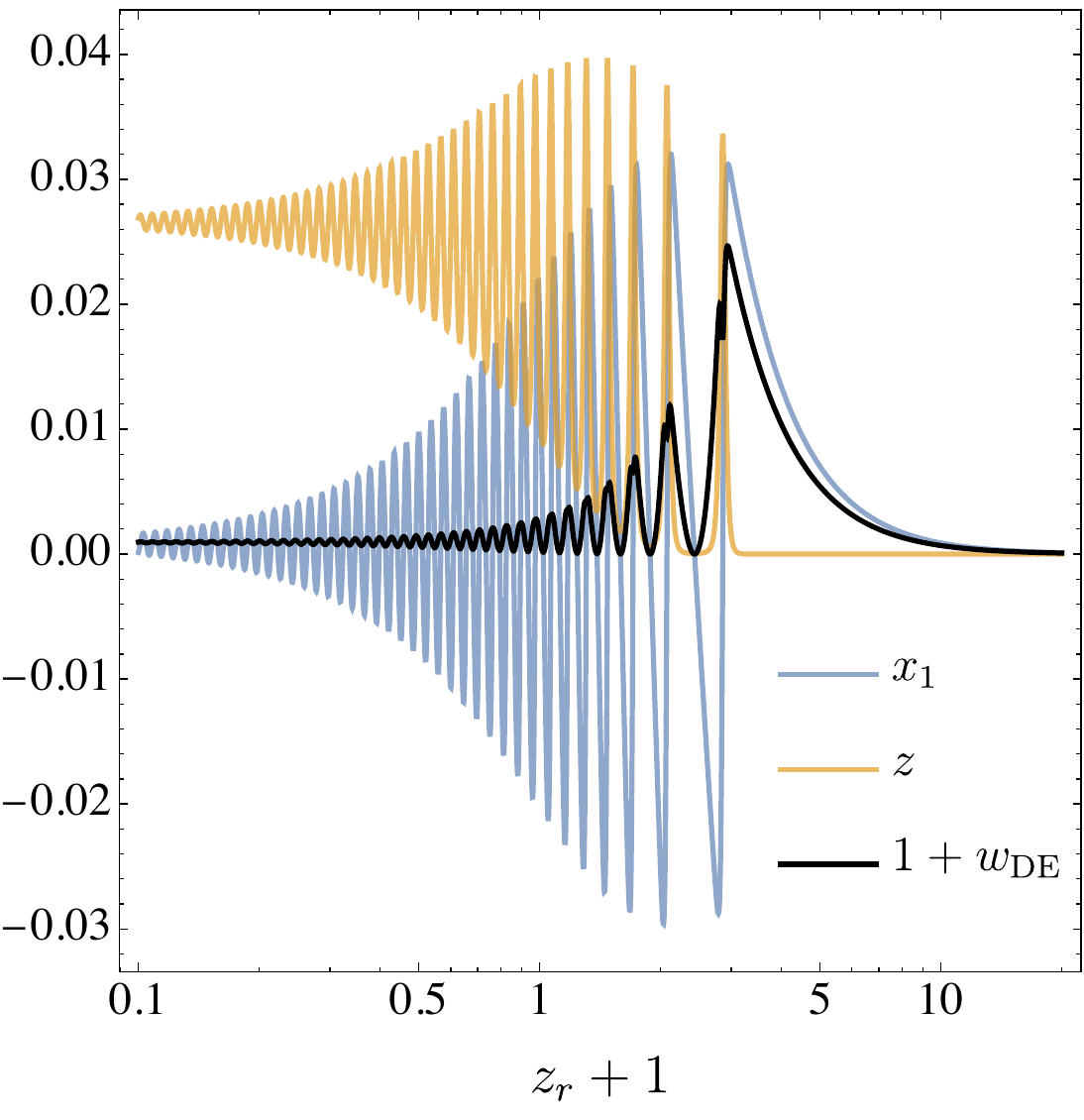}
\caption{(Left) Evolution of several cosmological parameters. The initial conditions were chosen deep in the radiation era. We see that the model can reproduce a viable expansion history, i.e., radiation dominance at early times (red dotted line), then a matter dominance (light brown dashed line), and dark energy dominance at late times (black solid line) characterized by $w_\text{eff} \simeq -1$ (blued dot-dashed line). Note that a early times, $w_\text{eff}$ behave as a stiff fluid and then as a ultra-relativistic fluid without relevant contributions to the energy budget. (Right) Evolution of the variables associated to the kinetic energies of the scalar field $\phi_1$ and the vector field $A$. The equation of state of dark energy $w_\text{DE}$ inherits the oscillations of these fields.}
\label{Fig: Evolution DE1}
\end{figure}

In the left panel of Figure~\ref{Fig: Evolution DE1}, we plot the evolution of $\Omega_r$, $\Omega_m$, and $\Omega_\text{DE} \equiv \rho_\text{DE}/(3M_\text{P}^2 H^2)$, and the effective equation of state. In particular it shows that the model evolves following the standard expansion history of the Universe: after the radiation dominated era ($\Omega_r \approx 1$ and  $w_\text{eff} \approx 1/3$) there is a radiation–matter transition around $z_r = 3000$. The matter-dominated epoch ($\Omega_m \approx 1$ and  $w_\text{eff} \approx 0$) then extends until $z_r \sim 0.3$ when the matter-dark energy transition takes place.
Importantly, during the radiation and matter-dominated epochs, the contribution of dark energy is subdominant. Indeed, we verified that $\Omega_\text{DE}$ follows the BBN constraint, $\Omega_\text{DE} < 0.045$, and the CMB constraint, $\Omega_\text{DE} < 0.02$ at $z_r = 50$. Dark energy dominance ($\Omega_\text{DE} \approx 1$ and $w_\text{eff} < -1/3$ for $z \rightarrow -1$) is characterised by a quasi-de Sitter evolution where $w_\text{DE} \approx -1$ and small oscillations at late times. As shown in Ref.~\cite{Orjuela-Quintana:2021zoe}, this oscillations are supported by the presence of the vector field. In the right panel of Figure~\ref{Fig: Evolution DE1}, we can see that, at late times, the kinetic energy of the scalar field, driving the accelerated expansion, and the kinetic energy of the vector field are of the same order, i.e., $x_1 \sim z$, and both oscillate. These oscillations directly influence the equation of state of dark energy. However, they remain relatively small and are likely beyond the observational scope of current and forthcoming surveys. See Ref.~\cite{Orjuela-Quintana:2021zoe} for further details about the cosmology of this fixed point.

We verified that indeed (DE1) is the attractor of the system by computing the values of the variables in the future, i.e., when $z_r \rightarrow - 1$ which we take as $N = 50$. These are given by
\begin{equation*}
x_1 \approx 8.2 \times 10^{-4}\,, \qquad x_2 \approx 1.5 \times 10^{-44}\,, \qquad y \approx 0.999\,,
\end{equation*}
\begin{equation}
z \approx 0.035\,, \qquad \Sigma \approx 8.2 \times 10^{-4}\,, \qquad w_\text{DE} \approx -0.998\,,
\end{equation}
which are consistent with the values computed from Table~\ref{tab:fpred}. In summary, the cosmological trajectory of the model compatible with the requirements from string theory, i.e., $\lambda = \sqrt{2p}$ and $\nu = \sqrt{2/p}$, is 
\begin{center}
Radiation \ $\rightarrow$ \ Matter \ $\rightarrow$ \ Anisotropic dark energy (DE1).
\end{center}
Note that although $x_2 = 0$ in the fixed point, the scalar field $\phi_2$ could be relevant to the cosmological dynamics well before the attractor is reached. However, we confirmed that the dynamics of the field $\phi_2$ is many orders of magnitude subdominant during the whole expansion history.

Attempting to implement a numerical solution for (DE2) poses a challenging task. The constrained parameter space necessary for it to function as an attractor point results in an exceedingly narrow region in phase space with trajectories rapidly converging to this point. To be more precise, the coexistence with the significantly more favoured (DE1) point implies that, despite the instability of (DE1) under the chosen parameters, it behaves akin to a pseudo-stable fixed point and trajectories spend virtually infinite time in (DE1) before transitioning to (DE2).

Another way to understand the unlikely nature of (DE2) comes directly from the solution of the equations of motion which for $\phi_2$ and $A^\mu$ read
\begin{equation}
\nabla_\mu \left( f^2(\phi_2) \nabla^\mu \phi_2 \right) = 0\,, \qquad  \nabla_\mu \left( h^2(\phi_1) F^{\mu \nu} \right) = 0\,.
\end{equation}
Indeed, for homogeneous fields in the context of the Bianchi I spacetime and a vector field as given in Eq.~\eqref{Eq: Bianchi metric},  yields the solutions:
\begin{equation}
\dot{\phi}_2 = \frac{c_{\phi_2}}{a^3 f^2(\phi_1)}\,, \qquad \dot{A} = \frac{c_A e^{-4\sigma}}{a h^2(\phi_1)}\,,
\end{equation}
with $c_{\phi_2}$ and $c_A$ integration constants. Consequently, the corresponding density contributions of these fields to the first Friedman equation are given by:
\begin{equation}
\rho_{\phi_2} \equiv \frac{1}{2} f^2(\phi_1) \dot{\phi}^2_2 \propto \frac{1}{a^6 f^2(\phi_1)}\,, \qquad \rho_A \equiv \frac{1}{2} h^2(\phi_1) \frac{\dot{A}^2 e^{4\sigma}}{a^2} \propto \frac{1}{a^4 h^2(\phi_1)}\,.
\end{equation}
At late time the systems approaches the fixed point and $\dot{\phi}_1 \approx 0$, thus the scalar field $\phi_1$ reaches its asymptotic value and the coupling functions $f(\phi_1)$ and $h(\phi_1)$ also stabilise at constant values. Therefore, it can be concluded that $\rho_{\phi_2} \propto 1/a^6$ while $\rho_A \propto 1/a^4$. Considering the rapid decay of $\rho_{\phi_2}$, two main observations can be made: $i)$ scalar fields can exhibit stiff matter behaviour, dominating the Universe's energy budget at very early times before the radiation-dominated epoch as shown in the left panel of Figure~\ref{Fig: Evolution DE1}, and $ii)$ at late times the contribution of the vector field to the total density consistently supersedes that of the scalar $\phi_2$. Hence, (DE1) describes the most natural asymptotic evolution of the system, while (DE2) corresponds only to a marginal option needing some degree of fine-tuning for its viability.

\subsection{Vector field from supergravity}
We have already mentioned the possibility of including some dynamics from gauge symmetries when contemplating a potential contribution to the scalar dynamics from the D-term scalar potential. Indeed, it is expected to have such, and to be general enough, we might also include the dynamics of the vector fields themselves. As before, we consider only the simplest case of a $U(1)$ sector, different from the one already included for the D-term scalar potential, as a non-vanishing D-term contribution implies a gauge symmetry breaking for the corresponding sector.
\\
The minimal setup will include a pure gauge $U(1)$, which can be justified by the absence of charged particles in the effective theory; that is, the charged sector turns out to be heavy compared with the energy scale involved in our studies. The general pure gauge sector in SUGRA is given by \cite{freedmanSupergravity2012}
\be 
\mathcal{L}_\text{gau}=\sqrt{-g}\left(-\frac14 \text{Re}(f_{AB})F^A_{\mu\nu}F^{\mu\nu~B}+\frac{\rmi}{4} \text{Im}(f_{AB})F^A_{\mu\nu}\tilde F^{\mu\nu\,B}\right)\,,
\ee
where the gauge kinetic function $f_{AB}$ is a holomorphic function of the chiral superfields, and $F_{\mu\nu}$ and $\tilde F_{\mu\nu}$ are the usual field strength and its dual. The main outshot extracted from this structure is the possibility of having field-dependent gauge couplings, once the gauge kinetic function is not trivial, and a clear connection between the kinetics term and the \emph{theta term}, i.e., the gauge coupling and the axion are superpartners. As before, we consider a diagonal metric in the gauge sector, $f_{AB}\propto \delta_{AB}$. Moreover, for homogenous field configurations, like the ones we are dealing with, the pseudoscalar term vanishes. Then we end up with a Lagrangian of the form
\be 
{\cal L}_\text{gau}=-\sqrt{-g}\frac14 \text{Re}(f(\Phi^I))F_{\mu\nu}F^{\mu\nu}\,.
\ee
As advertised when dealing with the D-term scalar dynamics, gauge kinetic functions in superstring compactifications are naturally linear in the moduli fields, i.e., \(f=\Psi/M_P\). Thus, once the canonical normalization is done, \(s=M_\text{P} e^{\sqrt{2/p}\phi_1/M_\text{P}}\).
\be \label{eq:LGauKin}
{\cal L}_\text{gau}=-\sqrt{-g}\frac14 e^{\sqrt{2/p}\phi_1/M_\text{P}}F_{\mu\nu}F^{\mu\nu}\,.
\ee
Therefore, an exponential dependence on the scalar field \(\phi_1\) is inherent in an effective field theory originating from string theory, providing a rationale for its inclusion in the preceding analysis. However, aligning the parameter \(\mu\) with the microscopic result yields \(\mu=-\sqrt{2/p}=-\nu\). Unfortunately, this falls beyond the attractor region for both fixed points, (DE1) and (DE2): firstly, as this requires positive values for \(\mu\); additionally, all parameters are anticipated to be of the same order, whereas an accelerated expansion scenario necessitates substantially small values of \(\lambda\) compared to \(\mu\) and \(\nu\) in each case; finally, achieving the relation \(\mu\approx 2\nu/3\) appears challenging given the precise expression from string compactifications.

Unfortunately, this non-isotropic backgrounds have not been the primary focus of investigations in superstring constructions. However, during the 1990s, several works emerged that delved into the direct solution of the equations of motion of the superstring sigma model with background fields \cite{callanStringsBackgroundFields1985,polchinskiStringTheoryVolume1998}. These endeavors sought to address the question of what types of solutions are consistent with non-zero background fields in the vacuum.
\\
Probably the first exploration of such a possibility was the work by Kaloper \cite{kaloperClosedBianchiUniverse1993}, where anisotropic spaces with a constant dilaton were discovered, albeit exhibiting a recollapse cosmology. Subsequently, Batakis \cite{batakisClassificationSpatiallyHomogeneous1995,batakisAnisotropicSpacetimesHomogeneous1995,batakisNewClassSpatially1995,batakisBianchitypeStringCosmology1995} classified possible homogeneous solutions that included a non-trivial 3-form field ($H$ field). These identified a specific class representing spaces with 4 non-compact dimensions akin to Bianchi spaces. Similar studies and results can be found in \cite{gasperiniHomogeneousConformalString1995,barrowSpatiallyHomogeneousString1997}, and \cite{copelandStringCosmologyTimedependent1995}. Notably, the latter work, in particular, incorporated a background with $B$ field, $H=\text{d}B$, and no $H$ as a background. Interestingly, it found that under these circumstances, an expanding universe is only possible in the anisotropic case. The inclusion of non-trivial moduli also presents solutions with this characteristic, as found in \cite{Barrow:1996gx,kawaiNonsingularBianchiType1999}. Non-isotropic solutions to the beta functions, now including 2-loop $\alpha'$ corrections and cases with central charge deficit, were also found recently in \cite{naderiAnisotropicHomogeneousString2017,naderiNoncriticalAnisotropicBianchi2018}, confirming the existence of non-isotropic solutions.

\section{Conclusions}

Our study delved into the cosmological implications of a simple yet versatile string-inspired model for dynamical dark energy. The model, rooted in a chiral modulus, exhibits common features in superstring compactifications, including a logarithmic Kähler potential, a modulus-independent superpotential, and a modulus-dependent gauge kinetic function in a pure gauge sector. This formulation results in a 4D model featuring two scalar fields and an abelian $U(1)$ vector field, coupled through non-trivial kinetic functions, and governed by a scalar potential — all displaying an exponential dependence on only one of the fields. These choices prompted a comprehensive three-parameter dynamical analysis. The study reviews previous results in which one of the fields plays a spectator role, and explore a novel situation where the interplay of all three fields is in the game.

In the scenario where the vector field assumes the role of a spectator, we observe a reduction to the case studied in Ref.~\cite{Cicoli:2020cfj}. Here, the interplay of two fields with non-geodesic trajectories in the scalar manifold gives rise to a quintessence attractor point, characterized by a steep potential. However, achieving correct values for cosmological parameters necessitates the coupling parameters to lie outside the preferred values derived from superstring string theory constructions.

The introduction of a non-trivial contribution from the vector field inherently results in a preferred spatial direction, prompting the consideration of a Bianchi-I background space-time. Although such a structural choice for non-compact dimensions is unconventional in string constructions, positive outcomes from direct examinations of string sigma models suggest not only its potential relevance but also its potential natural appearance in superstring compactifications. In this context, the model manifests a quintessence attractor fixed point, that we called (DE1), where the axionic scalar component assumes a trivial value—a scenario previously investigated by two of the authors in Ref.~\cite{Orjuela-Quintana:2021zoe}. In this particular instance, achieving alignment with the observed values and bounds for the equation of state parameters and shear necessitates a coupling with the vector field significantly stronger than the dynamics originating from the scalar potential.

This intriguing result aligns seamlessly with the swampland condition, as the dynamics from the vector field facilitate a slow-roll dynamics along a geodesic trajectory for the scalar field, despite its steep potential. In a broader context, this outcome extends the well-known results from non-geodesic trajectories in the scalar manifold, as formally discussed in Refs.~\cite{Achucarro:2018vey,Akrami:2020zfz}, and has already found application in various multifield scenarios.

Under these circumstances we analyse the cosmic evolution, which can be summarized as follows. An initially isotropic radiation domination epoch is followed by a matter domination period. Then, late accelerated expansion is driven by dark energy which also source anisotropy within current observational bounds~\cite{Campanelli:2010zx, Amirhashchi:2018nxl}. Another key feature of our dark energy model is the ocillating behavior of its equation of state. This feature has been found in other models~\cite{BeltranAlmeida:2019fou} and thoroughly explained in~\cite{Orjuela-Quintana:2021zoe}.

A second and novel fixed point (DE2) is also under consideration, where all three fields actively contribute to the dynamics. Remarkably, the system naturally accommodates this configuration, and the attractor quintessence point is realised for parameter values that are concurrently consistent with a small anisotropy. Intriguingly, the region hosting the attractor dynamics, marked also by accelerated behaviour, resides within a parameter space where the dynamics of the scalar fields are finely tuned to that of the vector field. However, this fine-tuning implies this attractor is significantly more constrained compared to the parameter space for the fixed point (DE1). In other words, it represents a much less probable scenario compared to the fixed point discussed above as can be seen in Figure~\ref{Fig: Bifurcation Curve}. We numerically corroborated that cosmological dynamics stay around (DE1), even in the case when the parameters $\lambda$, $\nu$, and $\mu$ are choice in the attraction region of (DE2), which is all around the plane $\mu = 2\nu/3$.

Unfortunately, in all the scenarios presented by the system, the parameter space required, either from stability requirements or cosmological observations, falls outside the suggested range from superstring compactifications. Specifically, in all four cases (two isotropic and two anisotropic attractors), a mandatory hierarchy for the coupling parameters is essential to obtain the correct value for the effective equation of state parameter and demonstrate a late accelerated epoch. In contrast, string compactifications naturally yield values of the same order. Additionally, for (DE1) and (DE2), achieving an attractor nature is only possible if the exponentials related to the kinetic coupling have the same sign in their exponents, whereas string compactifications are likely to produce exponents with opposite signs. Therefore, our findings intensify concerns about a top-down approach to constructing cosmological models for dark energy \cite{Cicoli:2021fsd,Brinkmann:2022oxy}. The simplest realisations struggle to replicate a viable expansion history, and even more complex models, though anticipated from a UV completion theory, might face similar challenges. However, it could be worth exploring configurations involving non-abelian gauge sectors, where input from the vector field dynamics might aid in realizing an accelerated late epoch.

To assess the generalizability of our findings, it is imperative to consider various corrections. These encompass quantum corrections to the Kähler potential (e.g., \cite{Becker:2002nn,Berg:2005ja}), non-perturbative deformations of the superpotential (e.g., \cite{Dine:1985rz, wittenDimensionalReductionSuperstring1985,Berg:2004ek}), quantum corrections to the effective Lagrangian following Supersymmetry breaking \cite{Coleman:1973jx}, and thermal corrections (e.g., \cite{Dolan:1973qd, Nakayama:2008ks,Anguelova:2009ht,Gallego:2020vbe}). These factors have the potential to significantly influence the dynamics of the system. Despite the prospect of these modifications, we maintain our anticipation that our primary concern regarding the improbable region in parameter space holds true on general grounds.

\acknowledgments
DG would like to thank Michele Cicoli and Francisco Pedro for our helpful email exchanges. Also to express his gratitude to the Physics Department of Universidad del Valle for their hospitality during the completion of this work. We are grateful to Michele Cicoli for their invaluable feedback on the near-final version of this manuscript.

\appendix

\section{String 4D effective actions}\label{app:4Dstringmodels}

The primary challenge in establishing a connection between superstring theory and reality lies in the complexity arising when transitioning from a ten-dimensional description to an effective four-dimensional theory. Although there exist only five consistent descriptions in ten dimensions, obtaining an effective four-dimensional theory leads to an exceedingly diverse situation. For instance, in the context of type II flux compactifications, the estimated number of 4D vacua is on the order of $10^{500}$ \cite{blumenhagenStatisticsSupersymmetricDbrane2005}. In such circumstances, formulating general statements applicable to model building seems nearly impossible.
\\
Nevertheless, the insights provided by four-dimensional ${\cal N}=1$ supersymmetry allow for the extraction of valuable information from the universal sector of the ten-dimensional effective theory. This approach was pioneered by Witten in the realm of Heterotic string compactifications \cite{wittenDimensionalReductionSuperstring1985} (see also \cite{fontSupersymmetryBreakingDuality1990a}). Supersymmetry dictates that the low-energy spectrum must be invariant under translations and $SU(3)$ rotations of the six compact dimensions.

The  starting point is the bosonic ten-dimensional action \cite{ibanezStringTheoryParticle2012}, working with dimensionless fields:
\be
S_{10} = \frac{1}{2\kappa_{10}^2}\int \text{d}^{10}x \sqrt{-G}e^{-2\phi}\left(R-2(\partial\phi)^2-\frac12 |H_3|^2-\frac{\alpha'}{4}\text{tr}_{v}|F|^{2}\right)\,,
\ee
where $\sqrt{\alpha'}=\ell_s$ denotes the fundamental string length controlling the world sheet perturbative expansion, and $\kappa_{10}=\frac12(2\pi)^7 \alpha'^4$ represents the reduced ten-dimensional Planck mass.
\\
Following the described procedure and disregarding the gauge sector, the dimensional reduction yields:
\bea\label{eq:4DbosonicHetdimred}
S_{4} &=& \frac{1}{2 M_\text{P}^2}\int \text{d}^4x \,\sqrt{-g}\left[R-2\partial_\mu\phi_4\partial^{\mu}\phi_4-\frac12e^{-4\phi_4}|H_3|^2\right.\cr
&&\qquad\qquad\qquad\left.-\frac12 \tilde g^{i\bar j}g^{k\bar \ell}\left(\partial_\mu\tilde g_{i\bar \ell}\partial^{\mu}\tilde g_{\bar jk}+\partial_\mu\tilde B_{i\bar \ell}\partial^{\mu}\tilde B_{\bar jk}\right)\right]\,,
\eea
where, $\kappa_{4}^2\equiv\kappa_{10}^2e^{2\phi}/V_6$ represents the 4D reduced Planck mass, $M_\text{P}^2=1/\kappa_4$, and
\be
\phi_4=\phi-\frac12\ln(V_6/\alpha'^3)\,,
\ee
where $V_6$ is the compact manifold volume, $H_3=\text{d}B_2$, and $\tilde g$ is the compact manifold metric, assumed to be a Calabi-Yau manifold, thus denoted as a 6D complex manifold with indices $i=1,\,2,\,3$. In four dimensions, the dual description of the $B_2$ field is achieved by a real scalar field:
\be
H^{\mu\nu\rho}=-\frac{1}{\sqrt{-g}}e^{4\phi_4}\epsilon^{\mu\nu\rho\tau}\partial_\tau\sigma\,.
\ee
Defining chiral superfields with lower components:
\be
S\equiv e^{-2\phi_4}+\rmi \sigma
\ee
and
\be
T_{i\bar j}\equiv \tilde g_{i\bar j}+\rmi B_{i\bar j}\,,
\ee
the non-gravitational part in \eqref{eq:4DbosonicHetdimred} can be expressed as the kinetic component of a ${\cal N}=1$ SUGRA theory. The Kähler potential is given by:
\be
\kappa_4K=-\ln(S+\bar S)-\ln\left[det(T+\bar T)\right]
\ee
The fields $T_{i\bar j}$ are the Kähler moduli, distinctly associated with the geometry of the internal dimensions. A specific scenario arises when these are diagonal, and all terms are equal, i.e., $T_{i\bar j}=T\delta_{i\bar j}$, leading to:
\be \label{eq:STKahlerPot}
\kappa_4K=-\ln(S+\bar S)-3\ln(T+\bar T)\,.
\ee
In the context of string theory, where the only free parameter is the string length, gauge coupling is expected to emerge as expectation values of functions of the fields. This is reflected in field-dependent kinetic functions for the gauge fields.
In the context of heterotic string theory, the gauge group is fixed in ten dimensions, explicitly either as $SO(32)$ or $E_8\times E_8$. The latter holds great appeal as it encompasses the typical grand unification groups, namely, $SO(10)$ and $SU(5)$. Generally, compactification to four dimensions involves the breaking of gauge symmetries, resulting, in general, in additional Abelian factors, in addition to any groups contained within the Standard Model. Explicit examples are presented in \cite{buchmullerSupersymmetricStandardModel2006,lebedevHeteroticMinilandscapeII2008,lebedevHeteroticRoadMSSM2008}. In this case, the coupling constant is universal and related to the dilaton, $\phi_4$, as evident from the gauge term in the action:
\begin{equation}
\int d^4x\sqrt{-g}\left(-\frac{1}{2}e^{-2\phi_4}\delta_{AB}F_{\mu\nu}^A F^{\mu\nu\,B}-\frac{\rmi}{2} a\delta_{AB}F_{\mu\nu}^A \tilde F^{\mu\nu\,B}\right)\,,
\end{equation}
such that the gauge kinetic function is given by
\be  
f_{AB}=S\delta_{AB}\,.
\ee
Trying to reproduce our setup in a heterotic scenario the pure $U(1)$ sector can be easily obtained from the hidden $E_8$ sector of the 10D theory after the compactifications.
Let us now examine the Kähler sector in Calabi-Yau compactifications. As mentioned earlier, Kähler moduli are linked to the geometric properties of the compact manifold, specifically to the volumes of four-cycles denoted as $\tau^a$. These volumes are defined through the Calabi-Yau manifold volume, $\V$, which is, in turn, a homogeneous function of degree 3 in the volumes of the two-cycles, $t_a$. Thus, $\tau^a = \V_{t_a}$, and the relation is expressed as:
\be
\label{eq:CYvolume}
t_a \tau^a = 3 \V \, . 
\ee 
Type IIB compactifications exhibit an approximate no-scale symmetry in the Kähler sector, albeit broken by $\alpha'$ corrections \cite{Becker:2002nn}. This implies that the associated Kähler potential becomes independent of those for other fields, leading to the relation:
\be 
K_{I}K^{I\bar J}K_{\bar J}=3\,.
\ee
Consequently, this results in a Kähler potential given by:
\be
\label{eq:KK}
\kappa_4K= -2 \ln  \V \, .
\ee
Subsequently, we recover the Kähler moduli part in \eqref{eq:STKahlerPot} with $\V=\tau^{3/2}=(T+\bar T)^{3/2}$. While neglecting the $\alpha'$ corrections can be initially justified in a large volume approximation, mixings with the axio-dilaton are expected in other scenarios.
\\
Moving on to gauge dynamics in type II superstrings, they are encoded either in D-branes or the Ramond-Ramond sector. For our focus, we consider the former case. The dimensional reduction of the Dp-brane action combines a Dirac-Born-Infield contribution:
\be\label{eq:DBIDp}
S_{Dp,DBI}=-\mu_p\int_{Dp}e^{-\phi}\sqrt{\det(G+B-2\pi \alpha'F)}\,,
\ee
with the brane tension given by $\mu_P=(\alpha')^{-(p+1)/2}/(2\pi)^p$, and a Chern-Simons term:
\be 
S_{Dp,SC}=\mu_p\int_{Dp}\sum_q C_q \text{tr} \, e^{2\pi\alpha' F-B}\,.
\ee
The involved fields are the pullbacks of the metric, $G$, and NSNS 2-form, $B$, on the Dp-brane worldvolume, the $C^{(n)}$ RR forms, and the worldvolume gauge field strengths, $F$. The expansion up to the second order in $F$ of the $DBI$ action leads to:
\be\label{eq:Dgauge4D}
S_{Dp,DBI}=\int d^4x\, \frac{(\alpha')^{(3-p)/2}}{(2\pi)^{p-2}}V_{\Sigma}e^{-\phi} F_{4d}^2+\cdots\,,
\ee
where $V_{\Sigma}=Vol(\Pi_{p-3})$ is the volume of the cycle $\Pi_{p-3}$ in the compact dimensions, with dimensionality $p-3$, allowing the brane to wrap around it. Hence, apart from the dilaton dependence, also encountered in the heterotic string case, the gauge couplings additionally depend on geometrical quantities:
\be 
\text{Re}(f)=\frac{(\alpha')^{(3-p)/2}}{(2\pi)^{p-2}}V_{\Sigma}e^{-\phi} \,.
\ee
The Chern-Simons action contributes to the imaginary part of the gauge kinetic function:
\be 
S_{Dp,SC}=\int_{\Pi_{p-3}}C_{p-3}\int d^4x \, \text{tr} \, (F\tilde F)\,.
\ee
The specific dependencies are situation-dependent, and explicit expressions are generally not feasible. However, in specific situations like compactifications on tori, it is possible to demonstrate, for instance, that the gauge kinetic function for D7-branes, appearing in type IIB string compactifications, on 4-cycles is given by:
\be 
2\pi f_i={\cal N}_i S+{\cal M}_i T_i\,,
\ee
with $T_i$ representing the Kähler moduli associated with the volume of the 4-cycle, and ${\cal N}$ and ${\cal M}$ being products of a pair of magnetic fluxes and a pair of wrapping numbers, i.e., products of integers, respectively. Therefore, both the axio-dilaton and the Kähler moduli generally appear in the gauge kinetic function in type IIB string compactifications, exhibiting a dependency similar to the one employed in the work. For more detailed descriptions and involved constructions with analogous results see for instance \cite{cremadesSUSYQuiversIntersecting2002,Blumenhagen:2003jy,lustScatteringGaugeMatter2004,grimmEffectiveActionCalabi2004,jockersEffectiveActionD7branes2005,fontSUSYbreakingSoftTerms2005,cicoliToricK3fibredCalabiYau2012a}.

Now, to ensure decoupling from the visible sector, it is imperative that the D-brane carrying our $U(1)$ field is positioned away from the branes supporting the standard model.

\bibliographystyle{JHEP}
\bibliography{Biblio.bib}

\providecommand{\href}[2]{#2}\begingroup\raggedright\begin{thebibliography}{100}

\bibitem{SupernovaSearchTeam:1998fmf}
{\scshape Supernova Search Team} collaboration, \emph{{Observational evidence
  from supernovae for an accelerating universe and a cosmological constant}},
  \href{https://doi.org/10.1086/300499}{\emph{Astron. J.} {\bfseries 116}
  (1998) 1009} [\href{https://arxiv.org/abs/astro-ph/9805201}{{\ttfamily
  astro-ph/9805201}}].

\bibitem{SupernovaCosmologyProject:1998vns}
{\scshape Supernova Cosmology Project} collaboration, \emph{{Measurements of
  $\Omega$ and $\Lambda$ from 42 high redshift supernovae}},
  \href{https://doi.org/10.1086/307221}{\emph{Astrophys. J.} {\bfseries 517}
  (1999) 565} [\href{https://arxiv.org/abs/astro-ph/9812133}{{\ttfamily
  astro-ph/9812133}}].

\bibitem{WMAP:2012fli}
{\scshape WMAP} collaboration, \emph{{Nine-Year Wilkinson Microwave Anisotropy
  Probe (WMAP) Observations: Final Maps and Results}},
  \href{https://doi.org/10.1088/0067-0049/208/2/20}{\emph{Astrophys. J. Suppl.}
  {\bfseries 208} (2013) 20} [\href{https://arxiv.org/abs/1212.5225}{{\ttfamily
  1212.5225}}].

\bibitem{Komatsu:2014ioa}
{\scshape WMAP Science Team} collaboration, \emph{{Results from the Wilkinson
  Microwave Anisotropy Probe}},
  \href{https://doi.org/10.1093/ptep/ptu083}{\emph{PTEP} {\bfseries 2014}
  (2014) 06B102} [\href{https://arxiv.org/abs/1404.5415}{{\ttfamily
  1404.5415}}].

\bibitem{Planck:2018vyg}
{\scshape Planck} collaboration, \emph{{Planck 2018 results. VI. Cosmological
  parameters}},
  \href{https://doi.org/10.1051/0004-6361/201833910}{\emph{Astron. Astrophys.}
  {\bfseries 641} (2020) A6}
  [\href{https://arxiv.org/abs/1807.06209}{{\ttfamily 1807.06209}}].

\bibitem{Agatsuma:2022ewd}
K.~Agatsuma, \emph{{Four hints and test candidates of the local cosmic
  expansion}}, \href{https://doi.org/10.1016/j.dark.2022.101134}{\emph{Phys.
  Dark Univ.} {\bfseries 38} (2022) 101134}
  [\href{https://arxiv.org/abs/2211.15668}{{\ttfamily 2211.15668}}].

\bibitem{Peebles:2002gy}
P.J.E.~Peebles and B.~Ratra, \emph{{The Cosmological Constant and Dark
  Energy}}, \href{https://doi.org/10.1103/RevModPhys.75.559}{\emph{Rev. Mod.
  Phys.} {\bfseries 75} (2003) 559}
  [\href{https://arxiv.org/abs/astro-ph/0207347}{{\ttfamily
  astro-ph/0207347}}].

\bibitem{Padmanabhan:2002ji}
T.~Padmanabhan, \emph{{Cosmological constant: The Weight of the vacuum}},
  \href{https://doi.org/10.1016/S0370-1573(03)00120-0}{\emph{Phys. Rept.}
  {\bfseries 380} (2003) 235}
  [\href{https://arxiv.org/abs/hep-th/0212290}{{\ttfamily hep-th/0212290}}].

\bibitem{Ostriker:1995rn}
J.P.~Ostriker and P.J.~Steinhardt, \emph{{Cosmic concordance}},
  \href{https://arxiv.org/abs/astro-ph/9505066}{{\ttfamily astro-ph/9505066}}.

\bibitem{Martin:2012bt}
J.~Martin, \emph{{Everything You Always Wanted To Know About The Cosmological
  Constant Problem (But Were Afraid To Ask)}},
  \href{https://doi.org/10.1016/j.crhy.2012.04.008}{\emph{Comptes Rendus
  Physique} {\bfseries 13} (2012) 566}
  [\href{https://arxiv.org/abs/1205.3365}{{\ttfamily 1205.3365}}].

\bibitem{Perivolaropoulos:2021jda}
L.~Perivolaropoulos and F.~Skara, \emph{{Challenges for
  \ensuremath{\Lambda}CDM: An update}},
  \href{https://doi.org/10.1016/j.newar.2022.101659}{\emph{New Astron. Rev.}
  {\bfseries 95} (2022) 101659}
  [\href{https://arxiv.org/abs/2105.05208}{{\ttfamily 2105.05208}}].

\bibitem{Riess:2016jrr}
A.G.~Riess et~al., \emph{{A 2.4\% Determination of the Local Value of the
  Hubble Constant}},
  \href{https://doi.org/10.3847/0004-637X/826/1/56}{\emph{Astrophys. J.}
  {\bfseries 826} (2016) 56}
  [\href{https://arxiv.org/abs/1604.01424}{{\ttfamily 1604.01424}}].

\bibitem{Riess:2020fzl}
A.G.~Riess, S.~Casertano, W.~Yuan, J.B.~Bowers, L.~Macri, J.C.~Zinn et~al.,
  \emph{{Cosmic Distances Calibrated to 1\% Precision with Gaia EDR3 Parallaxes
  and Hubble Space Telescope Photometry of 75 Milky Way Cepheids Confirm
  Tension with $\Lambda$CDM}},
  \href{https://doi.org/10.3847/2041-8213/abdbaf}{\emph{Astrophys. J. Lett.}
  {\bfseries 908} (2021) L6}
  [\href{https://arxiv.org/abs/2012.08534}{{\ttfamily 2012.08534}}].

\bibitem{Verde:2019ivm}
L.~Verde, T.~Treu and A.G.~Riess, \emph{{Tensions between the Early and the
  Late Universe}},
  \href{https://doi.org/10.1038/s41550-019-0902-0}{\emph{Nature Astron.}
  {\bfseries 3} (2019) 891} [\href{https://arxiv.org/abs/1907.10625}{{\ttfamily
  1907.10625}}].

\bibitem{Schoneberg:2021qvd}
N.~Sch\"oneberg, G.~Franco~Abell\'an, A.~P\'erez~S\'anchez, S.J.~Witte,
  V.~Poulin and J.~Lesgourgues, \emph{{The $H_0$ Olympics: A fair ranking of
  proposed models}},
  \href{https://doi.org/10.1016/j.physrep.2022.07.001}{\emph{Phys. Rept.}
  {\bfseries 984} (2022) 1} [\href{https://arxiv.org/abs/2107.10291}{{\ttfamily
  2107.10291}}].

\bibitem{DESI:2016fyo}
{\scshape DESI} collaboration, \emph{{The DESI Experiment Part I:
  Science,Targeting, and Survey Design}},
  \href{https://arxiv.org/abs/1611.00036}{{\ttfamily 1611.00036}}.

\bibitem{Amendola:2016saw}
L.~Amendola et~al., \emph{{Cosmology and fundamental physics with the Euclid
  satellite}}, \href{https://doi.org/10.1007/s41114-017-0010-3}{\emph{Living
  Rev. Rel.} {\bfseries 21} (2018) 2}
  [\href{https://arxiv.org/abs/1606.00180}{{\ttfamily 1606.00180}}].

\bibitem{Gebhardt:2021vfo}
K.~Gebhardt et~al., \emph{{The Hobby\textendash{}Eberly Telescope Dark Energy
  Experiment (HETDEX) Survey Design, Reductions, and Detections*}},
  \href{https://doi.org/10.3847/1538-4357/ac2e03}{\emph{Astrophys. J.}
  {\bfseries 923} (2021) 217}
  [\href{https://arxiv.org/abs/2110.04298}{{\ttfamily 2110.04298}}].

\bibitem{ratraCosmologicalConsequencesRolling1988}
B.~Ratra and P.J.E.~Peebles, \emph{Cosmological consequences of a rolling
  homogeneous scalar field},
  \href{https://doi.org/10.1103/PhysRevD.37.3406}{\emph{Phys. Rev. D}
  {\bfseries 37} (1988) 3406}.

\bibitem{Wetterich:1994bg}
C.~Wetterich, \emph{{The Cosmon model for an asymptotically vanishing time
  dependent cosmological ``constant''}}, {\emph{Astron. Astrophys.} {\bfseries
  301} (1995) 321} [\href{https://arxiv.org/abs/hep-th/9408025}{{\ttfamily
  hep-th/9408025}}].

\bibitem{Hellerman:2001yi}
S.~Hellerman, N.~Kaloper and L.~Susskind, \emph{{String theory and
  quintessence}},
  \href{https://doi.org/10.1088/1126-6708/2001/06/003}{\emph{JHEP} {\bfseries
  06} (2001) 003} [\href{https://arxiv.org/abs/hep-th/0104180}{{\ttfamily
  hep-th/0104180}}].

\bibitem{Fischler:2001yj}
W.~Fischler, A.~Kashani-Poor, R.~McNees and S.~Paban, \emph{{The Acceleration
  of the universe, a challenge for string theory}},
  \href{https://doi.org/10.1088/1126-6708/2001/07/003}{\emph{JHEP} {\bfseries
  07} (2001) 003} [\href{https://arxiv.org/abs/hep-th/0104181}{{\ttfamily
  hep-th/0104181}}].

\bibitem{Vafa:2005ui}
C.~Vafa, \emph{{The String landscape and the swampland}},
  \href{https://arxiv.org/abs/hep-th/0509212}{{\ttfamily hep-th/0509212}}.

\bibitem{Ooguri:2006in}
H.~Ooguri and C.~Vafa, \emph{{On the Geometry of the String Landscape and the
  Swampland}},
  \href{https://doi.org/10.1016/j.nuclphysb.2006.10.033}{\emph{Nucl. Phys. B}
  {\bfseries 766} (2007) 21}
  [\href{https://arxiv.org/abs/hep-th/0605264}{{\ttfamily hep-th/0605264}}].

\bibitem{Ooguri:2016pdq}
H.~Ooguri and C.~Vafa, \emph{{Non-supersymmetric AdS and the Swampland}},
  \href{https://doi.org/10.4310/ATMP.2017.v21.n7.a8}{\emph{Adv. Theor. Math.
  Phys.} {\bfseries 21} (2017) 1787}
  [\href{https://arxiv.org/abs/1610.01533}{{\ttfamily 1610.01533}}].

\bibitem{Brennan:2017rbf}
T.D.~Brennan, F.~Carta and C.~Vafa, \emph{{The String Landscape, the Swampland,
  and the Missing Corner}},
  \href{https://doi.org/10.22323/1.305.0015}{\emph{PoS} {\bfseries TASI2017}
  (2017) 015} [\href{https://arxiv.org/abs/1711.00864}{{\ttfamily
  1711.00864}}].

\bibitem{Palti:2019pca}
E.~Palti, \emph{{The Swampland: Introduction and Review}},
  \href{https://doi.org/10.1002/prop.201900037}{\emph{Fortsch. Phys.}
  {\bfseries 67} (2019) 1900037}
  [\href{https://arxiv.org/abs/1903.06239}{{\ttfamily 1903.06239}}].

\bibitem{VanRiet:2023pnx}
T.~Van~Riet and G.~Zoccarato, \emph{{Beginners lectures on flux
  compactifications and related Swampland topics}},
  \href{https://doi.org/10.1016/j.physrep.2023.11.003}{\emph{Phys. Rept.}
  {\bfseries 1049} (2024) 1}
  [\href{https://arxiv.org/abs/2305.01722}{{\ttfamily 2305.01722}}].

\bibitem{Brandenberger:2022sga}
R.~Brandenberger, \emph{{Perspectives on the Dark Sector}},  in \emph{{33rd
  Rencontres de Blois}: {Exploring the Dark Universe}}, 11, 2022
  [\href{https://arxiv.org/abs/2211.11273}{{\ttfamily 2211.11273}}].

\bibitem{Bergner:2001db}
Y.~Bergner and R.~Jackiw, \emph{{Integrable supersymmetric fluid mechanics from
  superstrings}},
  \href{https://doi.org/10.1016/S0375-9601(01)00305-X}{\emph{Phys. Lett. A}
  {\bfseries 284} (2001) 146}
  [\href{https://arxiv.org/abs/physics/0103092}{{\ttfamily physics/0103092}}].

\bibitem{Hassaine:2001is}
M.~Hassaine, \emph{{Supersymmetric Chaplygin gas}},
  \href{https://doi.org/10.1016/S0375-9601(01)00662-4}{\emph{Phys. Lett. A}
  {\bfseries 290} (2001) 157}
  [\href{https://arxiv.org/abs/hep-th/0106252}{{\ttfamily hep-th/0106252}}].

\bibitem{Bento:2002ps}
M.C.~Bento, O.~Bertolami and A.A.~Sen, \emph{{Generalized Chaplygin gas,
  accelerated expansion and dark energy matter unification}},
  \href{https://doi.org/10.1103/PhysRevD.66.043507}{\emph{Phys. Rev. D}
  {\bfseries 66} (2002) 043507}
  [\href{https://arxiv.org/abs/gr-qc/0202064}{{\ttfamily gr-qc/0202064}}].

\bibitem{Panda:2010uq}
S.~Panda, Y.~Sumitomo and S.P.~Trivedi, \emph{{Axions as Quintessence in String
  Theory}}, \href{https://doi.org/10.1103/PhysRevD.83.083506}{\emph{Phys. Rev.
  D} {\bfseries 83} (2011) 083506}
  [\href{https://arxiv.org/abs/1011.5877}{{\ttfamily 1011.5877}}].

\bibitem{Cicoli:2012tz}
M.~Cicoli, F.G.~Pedro and G.~Tasinato, \emph{{Natural Quintessence in String
  Theory}}, \href{https://doi.org/10.1088/1475-7516/2012/07/044}{\emph{JCAP}
  {\bfseries 07} (2012) 044} [\href{https://arxiv.org/abs/1203.6655}{{\ttfamily
  1203.6655}}].

\bibitem{DAmico:2018mnx}
G.~D'Amico, N.~Kaloper and A.~Lawrence, \emph{{Strongly Coupled Quintessence}},
  \href{https://doi.org/10.1103/PhysRevD.100.103504}{\emph{Phys. Rev. D}
  {\bfseries 100} (2019) 103504}
  [\href{https://arxiv.org/abs/1809.05109}{{\ttfamily 1809.05109}}].

\bibitem{Cicoli:2023qri}
M.~Cicoli, M.~Licheri, R.~Mahanta, E.~McDonough, F.G.~Pedro and M.~Scalisi,
  \emph{{Early Dark Energy in Type IIB String Theory}},
  \href{https://doi.org/10.1007/JHEP06(2023)052}{\emph{JHEP} {\bfseries 06}
  (2023) 052} [\href{https://arxiv.org/abs/2303.03414}{{\ttfamily
  2303.03414}}].

\bibitem{Kaloper:2005aj}
N.~Kaloper and L.~Sorbo, \emph{{Of pngb quintessence}},
  \href{https://doi.org/10.1088/1475-7516/2006/04/007}{\emph{JCAP} {\bfseries
  04} (2006) 007} [\href{https://arxiv.org/abs/astro-ph/0511543}{{\ttfamily
  astro-ph/0511543}}].

\bibitem{Kaloper:2008qs}
N.~Kaloper and L.~Sorbo, \emph{{Where in the String Landscape is
  Quintessence}}, \href{https://doi.org/10.1103/PhysRevD.79.043528}{\emph{Phys.
  Rev. D} {\bfseries 79} (2009) 043528}
  [\href{https://arxiv.org/abs/0810.5346}{{\ttfamily 0810.5346}}].

\bibitem{Blaback:2013fca}
J.~Bl\r{a}b\"ack, U.~Danielsson and G.~Dibitetto, \emph{{Accelerated Universes
  from type IIA Compactifications}},
  \href{https://doi.org/10.1088/1475-7516/2014/03/003}{\emph{JCAP} {\bfseries
  03} (2014) 003} [\href{https://arxiv.org/abs/1310.8300}{{\ttfamily
  1310.8300}}].

\bibitem{Kamionkowski:2014zda}
M.~Kamionkowski, J.~Pradler and D.G.E.~Walker, \emph{{Dark energy from the
  string axiverse}},
  \href{https://doi.org/10.1103/PhysRevLett.113.251302}{\emph{Phys. Rev. Lett.}
  {\bfseries 113} (2014) 251302}
  [\href{https://arxiv.org/abs/1409.0549}{{\ttfamily 1409.0549}}].

\bibitem{Alexander:2019rsc}
S.~Alexander and E.~McDonough, \emph{{Axion-Dilaton Destabilization and the
  Hubble Tension}},
  \href{https://doi.org/10.1016/j.physletb.2019.134830}{\emph{Phys. Lett. B}
  {\bfseries 797} (2019) 134830}
  [\href{https://arxiv.org/abs/1904.08912}{{\ttfamily 1904.08912}}].

\bibitem{McDonough:2022pku}
E.~McDonough and M.~Scalisi, \emph{{Towards Early Dark Energy in string
  theory}}, \href{https://doi.org/10.1007/JHEP10(2023)118}{\emph{JHEP}
  {\bfseries 10} (2023) 118}
  [\href{https://arxiv.org/abs/2209.00011}{{\ttfamily 2209.00011}}].

\bibitem{Achucarro:2018vey}
A.~Ach\'ucarro and G.A.~Palma, \emph{{The string swampland constraints require
  multi-field inflation}},
  \href{https://doi.org/10.1088/1475-7516/2019/02/041}{\emph{JCAP} {\bfseries
  02} (2019) 041} [\href{https://arxiv.org/abs/1807.04390}{{\ttfamily
  1807.04390}}].

\bibitem{Akrami:2020zfz}
Y.~Akrami, M.~Sasaki, A.R.~Solomon and V.~Vardanyan, \emph{{Multi-field dark
  energy: Cosmic acceleration on a steep potential}},
  \href{https://doi.org/10.1016/j.physletb.2021.136427}{\emph{Phys. Lett. B}
  {\bfseries 819} (2021) 136427}
  [\href{https://arxiv.org/abs/2008.13660}{{\ttfamily 2008.13660}}].

\bibitem{Brown:2017osf}
A.R.~Brown, \emph{{Hyperbolic Inflation}},
  \href{https://doi.org/10.1103/PhysRevLett.121.251601}{\emph{Phys. Rev. Lett.}
  {\bfseries 121} (2018) 251601}
  [\href{https://arxiv.org/abs/1705.03023}{{\ttfamily 1705.03023}}].

\bibitem{Christodoulidis:2019jsx}
P.~Christodoulidis, D.~Roest and E.I.~Sfakianakis, \emph{{Scaling attractors in
  multi-field inflation}},
  \href{https://doi.org/10.1088/1475-7516/2019/12/059}{\emph{JCAP} {\bfseries
  12} (2019) 059} [\href{https://arxiv.org/abs/1903.06116}{{\ttfamily
  1903.06116}}].

\bibitem{Bernardo:2022ztc}
H.~Bernardo, R.~Brandenberger and J.~Fr\"ohlich, \emph{{Towards a dark sector
  model from string theory}},
  \href{https://doi.org/10.1088/1475-7516/2022/09/040}{\emph{JCAP} {\bfseries
  09} (2022) 040} [\href{https://arxiv.org/abs/2201.04668}{{\ttfamily
  2201.04668}}].

\bibitem{Gasperini:2001pc}
M.~Gasperini, F.~Piazza and G.~Veneziano, \emph{{Quintessence as a runaway
  dilaton}}, \href{https://doi.org/10.1103/PhysRevD.65.023508}{\emph{Phys. Rev.
  D} {\bfseries 65} (2002) 023508}
  [\href{https://arxiv.org/abs/gr-qc/0108016}{{\ttfamily gr-qc/0108016}}].

\bibitem{vandeBruck:2019vzd}
C.~van~de Bruck and C.C.~Thomas, \emph{{Dark Energy, the Swampland and the
  Equivalence Principle}},
  \href{https://doi.org/10.1103/PhysRevD.100.023515}{\emph{Phys. Rev. D}
  {\bfseries 100} (2019) 023515}
  [\href{https://arxiv.org/abs/1904.07082}{{\ttfamily 1904.07082}}].

\bibitem{Gomes:2023dat}
J.M.~Gomes, E.~Hardy and S.~Parameswaran, \emph{{Dark Energy with a Little Help
  from its Friends}},  \href{https://arxiv.org/abs/2311.08888}{{\ttfamily
  2311.08888}}.

\bibitem{sonnerRecurrentAccelerationDilatonaxion2006}
J.~Sonner and P.K.~Townsend, \emph{{Recurrent acceleration in dilaton-axion
  cosmology}}, \href{https://doi.org/10.1103/PhysRevD.74.103508}{\emph{Phys.
  Rev. D} {\bfseries 74} (2006) 103508}
  [\href{https://arxiv.org/abs/hep-th/0608068}{{\ttfamily hep-th/0608068}}].

\bibitem{vandebruckQuintessenceDynamicsTwo2009}
C.~van~de Bruck and J.M.~Weller, \emph{{Quintessence dynamics with two scalar
  fields and mixed kinetic terms}},
  \href{https://doi.org/10.1103/PhysRevD.80.123014}{\emph{Phys. Rev. D}
  {\bfseries 80} (2009) 123014}
  [\href{https://arxiv.org/abs/0910.1934}{{\ttfamily 0910.1934}}].

\bibitem{Cicoli:2020cfj}
M.~Cicoli, G.~Dibitetto and F.G.~Pedro, \emph{{New accelerating solutions in
  late-time cosmology}},
  \href{https://doi.org/10.1103/PhysRevD.101.103524}{\emph{Phys. Rev. D}
  {\bfseries 101} (2020) 103524}
  [\href{https://arxiv.org/abs/2002.02695}{{\ttfamily 2002.02695}}].

\bibitem{cicoliOutSwamplandMultifield2020}
M.~Cicoli, G.~Dibitetto and F.G.~Pedro, \emph{{Out of the Swampland with
  Multifield Quintessence?}},
  \href{https://doi.org/10.1007/JHEP10(2020)035}{\emph{JHEP} {\bfseries 10}
  (2020) 035} [\href{https://arxiv.org/abs/2007.11011}{{\ttfamily
  2007.11011}}].

\bibitem{catenaAxionDilatonCosmology2008}
R.~Catena and J.~Moller, \emph{{Axion-dilaton cosmology and dark energy}},
  \href{https://doi.org/10.1088/1475-7516/2008/03/012}{\emph{JCAP} {\bfseries
  03} (2008) 012} [\href{https://arxiv.org/abs/0709.1931}{{\ttfamily
  0709.1931}}].

\bibitem{Heisenberg:2014rta}
L.~Heisenberg, \emph{{Generalization of the Proca Action}},
  \href{https://doi.org/10.1088/1475-7516/2014/05/015}{\emph{JCAP} {\bfseries
  05} (2014) 015} [\href{https://arxiv.org/abs/1402.7026}{{\ttfamily
  1402.7026}}].

\bibitem{Tasinato:2014eka}
G.~Tasinato, \emph{{Cosmic Acceleration from Abelian Symmetry Breaking}},
  \href{https://doi.org/10.1007/JHEP04(2014)067}{\emph{JHEP} {\bfseries 04}
  (2014) 067} [\href{https://arxiv.org/abs/1402.6450}{{\ttfamily 1402.6450}}].

\bibitem{BeltranJimenez:2016rff}
J.~Beltran~Jimenez and L.~Heisenberg, \emph{{Derivative self-interactions for a
  massive vector field}},
  \href{https://doi.org/10.1016/j.physletb.2016.04.017}{\emph{Phys. Lett. B}
  {\bfseries 757} (2016) 405}
  [\href{https://arxiv.org/abs/1602.03410}{{\ttfamily 1602.03410}}].

\bibitem{Allys:2015sht}
E.~Allys, P.~Peter and Y.~Rodriguez, \emph{{Generalized Proca action for an
  Abelian vector field}},
  \href{https://doi.org/10.1088/1475-7516/2016/02/004}{\emph{JCAP} {\bfseries
  02} (2016) 004} [\href{https://arxiv.org/abs/1511.03101}{{\ttfamily
  1511.03101}}].

\bibitem{Allys:2016jaq}
E.~Allys, J.P.~Beltran~Almeida, P.~Peter and Y.~Rodr\'\i{}guez, \emph{{On the
  4D generalized Proca action for an Abelian vector field}},
  \href{https://doi.org/10.1088/1475-7516/2016/09/026}{\emph{JCAP} {\bfseries
  09} (2016) 026} [\href{https://arxiv.org/abs/1605.08355}{{\ttfamily
  1605.08355}}].

\bibitem{DeFelice:2016yws}
A.~De~Felice, L.~Heisenberg, R.~Kase, S.~Mukohyama, S.~Tsujikawa and
  Y.-l.~Zhang, \emph{{Cosmology in generalized Proca theories}},
  \href{https://doi.org/10.1088/1475-7516/2016/06/048}{\emph{JCAP} {\bfseries
  06} (2016) 048} [\href{https://arxiv.org/abs/1603.05806}{{\ttfamily
  1603.05806}}].

\bibitem{Heisenberg:2018mxx}
L.~Heisenberg, R.~Kase and S.~Tsujikawa, \emph{{Cosmology in
  scalar-vector-tensor theories}},
  \href{https://doi.org/10.1103/PhysRevD.98.024038}{\emph{Phys. Rev. D}
  {\bfseries 98} (2018) 024038}
  [\href{https://arxiv.org/abs/1805.01066}{{\ttfamily 1805.01066}}].

\bibitem{Rinaldi:2015iza}
M.~Rinaldi, \emph{{Dark energy as a fixed point of the Einstein Yang-Mills
  Higgs Equations}},
  \href{https://doi.org/10.1088/1475-7516/2015/10/023}{\emph{JCAP} {\bfseries
  10} (2015) 023} [\href{https://arxiv.org/abs/1508.04576}{{\ttfamily
  1508.04576}}].

\bibitem{Alvarez:2019ues}
M.~\'Alvarez, J.B.~Orjuela-Quintana, Y.~Rodriguez and C.A.~Valenzuela-Toledo,
  \emph{{Einstein Yang\textendash{}Mills Higgs dark energy revisited}},
  \href{https://doi.org/10.1088/1361-6382/ab3775}{\emph{Class. Quant. Grav.}
  {\bfseries 36} (2019) 195004}
  [\href{https://arxiv.org/abs/1901.04624}{{\ttfamily 1901.04624}}].

\bibitem{Guarnizo:2020pkj}
A.~Guarnizo, J.B.~Orjuela-Quintana and C.A.~Valenzuela-Toledo, \emph{{Dynamical
  analysis of cosmological models with non-Abelian gauge vector fields}},
  \href{https://doi.org/10.1103/PhysRevD.102.083507}{\emph{Phys. Rev. D}
  {\bfseries 102} (2020) 083507}
  [\href{https://arxiv.org/abs/2007.12964}{{\ttfamily 2007.12964}}].

\bibitem{Armendariz-Picon:2004say}
C.~Armendariz-Picon, \emph{{Could dark energy be vector-like?}},
  \href{https://doi.org/10.1088/1475-7516/2004/07/007}{\emph{JCAP} {\bfseries
  07} (2004) 007} [\href{https://arxiv.org/abs/astro-ph/0405267}{{\ttfamily
  astro-ph/0405267}}].

\bibitem{Orjuela-Quintana:2020klr}
J.B.~Orjuela-Quintana, M.~Alvarez, C.A.~Valenzuela-Toledo and Y.~Rodriguez,
  \emph{{Anisotropic Einstein Yang-Mills Higgs Dark Energy}},
  \href{https://doi.org/10.1088/1475-7516/2020/10/019}{\emph{JCAP} {\bfseries
  10} (2020) 019} [\href{https://arxiv.org/abs/2006.14016}{{\ttfamily
  2006.14016}}].

\bibitem{Motoa-Manzano:2020mwe}
J.~Motoa-Manzano, J.~Bayron Orjuela-Quintana, T.S.~Pereira and
  C.A.~Valenzuela-Toledo, \emph{{Anisotropic solid dark energy}},
  \href{https://doi.org/10.1016/j.dark.2021.100806}{\emph{Phys. Dark Univ.}
  {\bfseries 32} (2021) 100806}
  [\href{https://arxiv.org/abs/2012.09946}{{\ttfamily 2012.09946}}].

\bibitem{Kase:2018nwt}
R.~Kase and S.~Tsujikawa, \emph{{Dark energy in scalar-vector-tensor
  theories}}, \href{https://doi.org/10.1088/1475-7516/2018/11/024}{\emph{JCAP}
  {\bfseries 11} (2018) 024}
  [\href{https://arxiv.org/abs/1805.11919}{{\ttfamily 1805.11919}}].

\bibitem{Cardona:2022lcz}
W.~Cardona, J.B.~Orjuela-Quintana and C.A.~Valenzuela-Toledo, \emph{{An
  effective fluid description of scalar-vector-tensor theories under the
  sub-horizon and quasi-static approximations}},
  \href{https://doi.org/10.1088/1475-7516/2022/08/059}{\emph{JCAP} {\bfseries
  08} (2022) 059} [\href{https://arxiv.org/abs/2206.02895}{{\ttfamily
  2206.02895}}].

\bibitem{Thorsrud:2012mu}
M.~Thorsrud, D.F.~Mota and S.~Hervik, \emph{{Cosmology of a Scalar Field
  Coupled to Matter and an Isotropy-Violating Maxwell Field}},
  \href{https://doi.org/10.1007/JHEP10(2012)066}{\emph{JHEP} {\bfseries 10}
  (2012) 066} [\href{https://arxiv.org/abs/1205.6261}{{\ttfamily 1205.6261}}].

\bibitem{Orjuela-Quintana:2021zoe}
J.B.~Orjuela-Quintana and C.A.~Valenzuela-Toledo, \emph{{Anisotropic
  k-essence}}, \href{https://doi.org/10.1016/j.dark.2021.100857}{\emph{Phys.
  Dark Univ.} {\bfseries 33} (2021) 100857}
  [\href{https://arxiv.org/abs/2106.06432}{{\ttfamily 2106.06432}}].

\bibitem{freedmanSupergravity2012}
D.Z.~Freedman and A.~Van~Proeyen, \emph{{Supergravity}}, Cambridge Univ. Press,
  Cambridge, UK (5, 2012),
  \href{https://doi.org/10.1017/CBO9781139026833}{10.1017/CBO9781139026833}.

\bibitem{Cicoli:2023opf}
M.~Cicoli, J.P.~Conlon, A.~Maharana, S.~Parameswaran, F.~Quevedo and I.~Zavala,
  \emph{{String Cosmology: from the Early Universe to Today}},
  \href{https://arxiv.org/abs/2303.04819}{{\ttfamily 2303.04819}}.

\bibitem{font1990supersymmetry}
A.~Font, L.E.~Ibanez, D.~Lust and F.~Quevedo, \emph{{Supersymmetry Breaking
  From Duality Invariant Gaugino Condensation}},
  \href{https://doi.org/10.1016/0370-2693(90)90665-S}{\emph{Phys. Lett. B}
  {\bfseries 245} (1990) 401}.

\bibitem{ferrara1990duality}
S.~Ferrara, N.~Magnoli, T.R.~Taylor and G.~Veneziano, \emph{{Duality and
  supersymmetry breaking in string theory}},
  \href{https://doi.org/10.1016/0370-2693(90)90666-T}{\emph{Phys. Lett. B}
  {\bfseries 245} (1990) 409}.

\bibitem{Kachru:2003aw}
S.~Kachru, R.~Kallosh, A.D.~Linde and S.P.~Trivedi, \emph{{De Sitter vacua in
  string theory}},
  \href{https://doi.org/10.1103/PhysRevD.68.046005}{\emph{Phys. Rev. D}
  {\bfseries 68} (2003) 046005}
  [\href{https://arxiv.org/abs/hep-th/0301240}{{\ttfamily hep-th/0301240}}].

\bibitem{Balasubramanian:2005zx}
V.~Balasubramanian, P.~Berglund, J.P.~Conlon and F.~Quevedo, \emph{{Systematics
  of moduli stabilisation in Calabi-Yau flux compactifications}},
  \href{https://doi.org/10.1088/1126-6708/2005/03/007}{\emph{JHEP} {\bfseries
  03} (2005) 007} [\href{https://arxiv.org/abs/hep-th/0502058}{{\ttfamily
  hep-th/0502058}}].

\bibitem{DeWolfe:2005uu}
O.~DeWolfe, A.~Giryavets, S.~Kachru and W.~Taylor, \emph{{Type IIA moduli
  stabilization}},
  \href{https://doi.org/10.1088/1126-6708/2005/07/066}{\emph{JHEP} {\bfseries
  07} (2005) 066} [\href{https://arxiv.org/abs/hep-th/0505160}{{\ttfamily
  hep-th/0505160}}].

\bibitem{Camara:2005dc}
P.G.~Camara, A.~Font and L.E.~Ibanez, \emph{{Fluxes, moduli fixing and
  MSSM-like vacua in a simple IIA orientifold}},
  \href{https://doi.org/10.1088/1126-6708/2005/09/013}{\emph{JHEP} {\bfseries
  09} (2005) 013} [\href{https://arxiv.org/abs/hep-th/0506066}{{\ttfamily
  hep-th/0506066}}].

\bibitem{Gallego:2017dvd}
D.~Gallego, M.C.D.~Marsh, B.~Vercnocke and T.~Wrase, \emph{{A New Class of de
  Sitter Vacua in Type IIB Large Volume Compactifications}},
  \href{https://doi.org/10.1007/JHEP10(2017)193}{\emph{JHEP} {\bfseries 10}
  (2017) 193} [\href{https://arxiv.org/abs/1707.01095}{{\ttfamily
  1707.01095}}].

\bibitem{McAllister:2023vgy}
L.~McAllister and F.~Quevedo, \emph{{Moduli Stabilization in String Theory}},
  \href{https://arxiv.org/abs/2310.20559}{{\ttfamily 2310.20559}}.

\bibitem{dineFayetIliopoulosTermsString1987}
M.~Dine, N.~Seiberg and E.~Witten, \emph{Fayet-{Iliopoulos} terms in string
  theory}, \href{https://doi.org/10.1016/0550-3213(87)90395-6}{\emph{Nuclear
  Physics B} {\bfseries 289} (1987) 589}.

\bibitem{dineTermsTermsString1987}
M.~Dine, I.~Ichinose and N.~Seiberg, \emph{F terms and {D} terms in string
  theory}, \href{https://doi.org/10.1016/0550-3213(87)90072-1}{\emph{Nuclear
  Physics B} {\bfseries 293} (1987) 253}.

\bibitem{atickStringCalculationFayetiliopoulos1987}
J.J.~Atick, L.J.~Dixon and A.~Sen, \emph{String calculation of fayet-iliopoulos
  {D}-terms in arbitrary supersymmetric compactifications},
  \href{https://doi.org/10.1016/0550-3213(87)90639-0}{\emph{Nuclear Physics B}
  {\bfseries 292} (1987) 109}.

\bibitem{gallegoEffectiveDescriptionLandscape2009}
D.~Gallego and M.~Serone, \emph{{An Effective Description of the Landscape -
  II}}, \href{https://doi.org/10.1088/1126-6708/2009/06/057}{\emph{JHEP}
  {\bfseries 06} (2009) 057} [\href{https://arxiv.org/abs/0904.2537}{{\ttfamily
  0904.2537}}].

\bibitem{Wainwright2009}
J.~Wainwright and G.F.R.~Ellis, \emph{Dynamical Systems in Cosmology},
  Cambridge University Press, New York (2009).

\bibitem{BeltranAlmeida:2019fou}
J.P.~Beltr\'an~Almeida, A.~Guarnizo, R.~Kase, S.~Tsujikawa and
  C.A.~Valenzuela-Toledo, \emph{{Anisotropic $2$-form dark energy}},
  \href{https://doi.org/10.1016/j.physletb.2019.05.008}{\emph{Phys. Lett. B}
  {\bfseries 793} (2019) 396}
  [\href{https://arxiv.org/abs/1902.05846}{{\ttfamily 1902.05846}}].

\bibitem{Koivisto:2008xf}
T.~Koivisto and D.F.~Mota, \emph{{Vector Field Models of Inflation and Dark
  Energy}}, \href{https://doi.org/10.1088/1475-7516/2008/08/021}{\emph{JCAP}
  {\bfseries 08} (2008) 021} [\href{https://arxiv.org/abs/0805.4229}{{\ttfamily
  0805.4229}}].

\bibitem{Secrest:2020has}
N.J.~Secrest, S.~von Hausegger, M.~Rameez, R.~Mohayaee, S.~Sarkar and J.~Colin,
  \emph{{A Test of the Cosmological Principle with Quasars}},
  \href{https://doi.org/10.3847/2041-8213/abdd40}{\emph{Astrophys. J. Lett.}
  {\bfseries 908} (2021) L51}
  [\href{https://arxiv.org/abs/2009.14826}{{\ttfamily 2009.14826}}].

\bibitem{Secrest:2022uvx}
N.J.~Secrest, S.~von Hausegger, M.~Rameez, R.~Mohayaee and S.~Sarkar, \emph{{A
  Challenge to the Standard Cosmological Model}},
  \href{https://doi.org/10.3847/2041-8213/ac88c0}{\emph{Astrophys. J. Lett.}
  {\bfseries 937} (2022) L31}
  [\href{https://arxiv.org/abs/2206.05624}{{\ttfamily 2206.05624}}].

\bibitem{Dam:2022wwh}
L.~Dam, G.F.~Lewis and B.J.~Brewer, \emph{{Testing the cosmological principle
  with CatWISE quasars: a bayesian analysis of the number-count dipole}},
  \href{https://doi.org/10.1093/mnras/stad2322}{\emph{Mon. Not. Roy. Astron.
  Soc.} {\bfseries 525} (2023) 231}
  [\href{https://arxiv.org/abs/2212.07733}{{\ttfamily 2212.07733}}].

\bibitem{Jones:2023ncn}
J.~Jones, C.J.~Copi, G.D.~Starkman and Y.~Akrami, \emph{{The Universe is not
  statistically isotropic}},
  \href{https://arxiv.org/abs/2310.12859}{{\ttfamily 2310.12859}}.

\bibitem{Karciauskas:2016pxn}
M.~Kar\v{c}iauskas, \emph{{Dynamical Analysis of Anisotropic Inflation}},
  \href{https://doi.org/10.1142/S0217732316400022}{\emph{Mod. Phys. Lett. A}
  {\bfseries 31} (2016) 1640002}
  [\href{https://arxiv.org/abs/1604.00269}{{\ttfamily 1604.00269}}].

\bibitem{Wald:1983ky}
R.M.~Wald, \emph{{Asymptotic behavior of homogeneous cosmological models in the
  presence of a positive cosmological constant}},
  \href{https://doi.org/10.1103/PhysRevD.28.2118}{\emph{Phys. Rev. D}
  {\bfseries 28} (1983) 2118}.

\bibitem{Campanelli:2010zx}
L.~Campanelli, P.~Cea, G.L.~Fogli and A.~Marrone, \emph{{Testing the Isotropy
  of the Universe with Type Ia Supernovae}},
  \href{https://doi.org/10.1103/PhysRevD.83.103503}{\emph{Phys. Rev. D}
  {\bfseries 83} (2011) 103503}
  [\href{https://arxiv.org/abs/1012.5596}{{\ttfamily 1012.5596}}].

\bibitem{Amirhashchi:2018nxl}
H.~Amirhashchi and S.~Amirhashchi, \emph{{Constraining Bianchi Type I Universe
  With Type Ia Supernova and H(z) Data}},
  \href{https://doi.org/10.1016/j.dark.2020.100557}{\emph{Phys. Dark Univ.}
  {\bfseries 29} (2020) 100557}
  [\href{https://arxiv.org/abs/1802.04251}{{\ttfamily 1802.04251}}].

\bibitem{Garcia-Serna:2023xfw}
S.~Garc\'\i{}a-Serna, J.B.~Orjuela-Quintana, C.A.~Valenzuela-Toledo and
  H.~Ocampo-Dur\'an, \emph{{Reconstructing the parameter space of nonanalytical
  cosmological fixed points}},
  \href{https://doi.org/10.1142/S0218271823500736}{\emph{Int. J. Mod. Phys. D}
  {\bfseries 32} (2023) 2350073}
  [\href{https://arxiv.org/abs/2302.09181}{{\ttfamily 2302.09181}}].

\bibitem{callanStringsBackgroundFields1985}
C.G.~Callan, D.~Friedan, E.J.~Martinec and M.J.~Perry, \emph{Strings in
  background fields},
  \href{https://doi.org/10.1016/0550-3213(85)90506-1}{\emph{Nuclear Physics B}
  {\bfseries 262} (1985) 593}.

\bibitem{polchinskiStringTheoryVolume1998}
J.~Polchinski, \emph{String {Theory}: {Volume} 1: {An} {Introduction} to the
  {Bosonic} {String}}, vol.~1 of \emph{Cambridge {Monographs} on {Mathematical}
  {Physics}}, Cambridge University Press, Cambridge (1998),
  \href{https://doi.org/10.1017/CBO9780511816079}{10.1017/CBO9780511816079}.

\bibitem{kaloperClosedBianchiUniverse1993}
N.~Kaloper, \emph{{The Closed Bianchi I universe in string theory}},
  \href{https://doi.org/10.1142/S021773239300043X}{\emph{Mod. Phys. Lett. A}
  {\bfseries 8} (1993) 421}
  [\href{https://arxiv.org/abs/hep-th/9208025}{{\ttfamily hep-th/9208025}}].

\bibitem{batakisClassificationSpatiallyHomogeneous1995}
N.A.~Batakis, \emph{{On the classification of spatially homogeneous 4-D string
  backgrounds}},  \href{https://arxiv.org/abs/hep-th/9502136}{{\ttfamily
  hep-th/9502136}}.

\bibitem{batakisAnisotropicSpacetimesHomogeneous1995}
N.A.~Batakis and A.A.~Kehagias, \emph{{Anisotropic space-times in homogeneous
  string cosmology}},
  \href{https://doi.org/10.1016/0550-3213(95)00249-R}{\emph{Nucl. Phys. B}
  {\bfseries 449} (1995) 248}
  [\href{https://arxiv.org/abs/hep-th/9502007}{{\ttfamily hep-th/9502007}}].

\bibitem{batakisNewClassSpatially1995}
N.A.~Batakis, \emph{{A New class of spatially homogeneous 4-D string
  backgrounds}},
  \href{https://doi.org/10.1016/0370-2693(95)00632-U}{\emph{Phys. Lett. B}
  {\bfseries 353} (1995) 450}
  [\href{https://arxiv.org/abs/hep-th/9503142}{{\ttfamily hep-th/9503142}}].

\bibitem{batakisBianchitypeStringCosmology1995}
N.A.~Batakis, \emph{Bianchi-type string cosmology},
  \href{https://doi.org/10.1016/0370-2693(95)00582-6}{\emph{Physics Letters B}
  {\bfseries 353} (1995) 39}.

\bibitem{gasperiniHomogeneousConformalString1995}
M.~Gasperini and R.~Ricci, \emph{{Homogeneous conformal string backgrounds}},
  \href{https://doi.org/10.1088/0264-9381/12/3/006}{\emph{Class. Quant. Grav.}
  {\bfseries 12} (1995) 677}
  [\href{https://arxiv.org/abs/hep-th/9501055}{{\ttfamily hep-th/9501055}}].

\bibitem{barrowSpatiallyHomogeneousString1997}
J.D.~Barrow and K.E.~Kunze, \emph{{Spatially homogeneous string cosmologies}},
  \href{https://doi.org/10.1103/PhysRevD.55.623}{\emph{Phys. Rev. D} {\bfseries
  55} (1997) 623} [\href{https://arxiv.org/abs/hep-th/9608045}{{\ttfamily
  hep-th/9608045}}].

\bibitem{copelandStringCosmologyTimedependent1995}
E.J.~Copeland, A.~Lahiri and D.~Wands, \emph{{String cosmology with a time
  dependent antisymmetric tensor potential}},
  \href{https://doi.org/10.1103/PhysRevD.51.1569}{\emph{Phys. Rev. D}
  {\bfseries 51} (1995) 1569}
  [\href{https://arxiv.org/abs/hep-th/9410136}{{\ttfamily hep-th/9410136}}].

\bibitem{Barrow:1996gx}
J.D.~Barrow and M.P.~Dabrowski, \emph{{Kantowski-Sachs string cosmologies}},
  \href{https://doi.org/10.1103/PhysRevD.55.630}{\emph{Phys. Rev. D} {\bfseries
  55} (1997) 630} [\href{https://arxiv.org/abs/hep-th/9608136}{{\ttfamily
  hep-th/9608136}}].

\bibitem{kawaiNonsingularBianchiType1999}
S.~Kawai and J.~Soda, \emph{{Nonsingular Bianchi type 1 cosmological solutions
  from 1 loop superstring effective action}},
  \href{https://doi.org/10.1103/PhysRevD.59.063506}{\emph{Phys. Rev. D}
  {\bfseries 59} (1999) 063506}
  [\href{https://arxiv.org/abs/gr-qc/9807060}{{\ttfamily gr-qc/9807060}}].

\bibitem{naderiAnisotropicHomogeneousString2017}
F.~Naderi and A.~Rezaei-Aghdam, \emph{Anisotropic homogeneous string cosmology
  with two-loop corrections},
  \href{https://doi.org/10.1016/j.nuclphysb.2017.08.005}{\emph{Nuclear Physics
  B} {\bfseries 923} (2017) 416}.

\bibitem{naderiNoncriticalAnisotropicBianchi2018}
F.~Naderi, A.~Rezaei-Aghdam and F.~Darabi, \emph{Non-critical anisotropic
  {Bianchi} type \${I}\$ string cosmology with
  \${\textbackslash}alpha'\$-corrections},
  \href{https://doi.org/10.1103/PhysRevD.98.026009}{\emph{Phys. Rev. D}
  {\bfseries 98} (2018) 026009}.

\bibitem{Cicoli:2021fsd}
M.~Cicoli, F.~Cunillera, A.~Padilla and F.G.~Pedro, \emph{{Quintessence and the
  Swampland: The Parametrically Controlled Regime of Moduli Space}},
  \href{https://doi.org/10.1002/prop.202200009}{\emph{Fortsch. Phys.}
  {\bfseries 70} (2022) 2200009}
  [\href{https://arxiv.org/abs/2112.10779}{{\ttfamily 2112.10779}}].

\bibitem{Brinkmann:2022oxy}
M.~Brinkmann, M.~Cicoli, G.~Dibitetto and F.G.~Pedro, \emph{{Stringy multifield
  quintessence and the Swampland}},
  \href{https://doi.org/10.1007/JHEP11(2022)044}{\emph{JHEP} {\bfseries 11}
  (2022) 044} [\href{https://arxiv.org/abs/2206.10649}{{\ttfamily
  2206.10649}}].

\bibitem{Becker:2002nn}
K.~Becker, M.~Becker, M.~Haack and J.~Louis, \emph{{Supersymmetry breaking and
  alpha-prime corrections to flux induced potentials}},
  \href{https://doi.org/10.1088/1126-6708/2002/06/060}{\emph{JHEP} {\bfseries
  06} (2002) 060} [\href{https://arxiv.org/abs/hep-th/0204254}{{\ttfamily
  hep-th/0204254}}].

\bibitem{Berg:2005ja}
M.~Berg, M.~Haack and B.~Kors, \emph{{String loop corrections to Kahler
  potentials in orientifolds}},
  \href{https://doi.org/10.1088/1126-6708/2005/11/030}{\emph{JHEP} {\bfseries
  11} (2005) 030} [\href{https://arxiv.org/abs/hep-th/0508043}{{\ttfamily
  hep-th/0508043}}].

\bibitem{Dine:1985rz}
M.~Dine, R.~Rohm, N.~Seiberg and E.~Witten, \emph{{Gluino Condensation in
  Superstring Models}},
  \href{https://doi.org/10.1016/0370-2693(85)91354-1}{\emph{Phys. Lett. B}
  {\bfseries 156} (1985) 55}.

\bibitem{wittenDimensionalReductionSuperstring1985}
E.~Witten, \emph{Dimensional reduction of superstring models},
  \href{https://doi.org/10.1016/0370-2693(85)90976-1}{\emph{Physics Letters B}
  {\bfseries 155} (1985) 151}.

\bibitem{Berg:2004ek}
M.~Berg, M.~Haack and B.~Kors, \emph{{Loop corrections to volume moduli and
  inflation in string theory}},
  \href{https://doi.org/10.1103/PhysRevD.71.026005}{\emph{Phys. Rev. D}
  {\bfseries 71} (2005) 026005}
  [\href{https://arxiv.org/abs/hep-th/0404087}{{\ttfamily hep-th/0404087}}].

\bibitem{Coleman:1973jx}
S.R.~Coleman and E.J.~Weinberg, \emph{{Radiative Corrections as the Origin of
  Spontaneous Symmetry Breaking}},
  \href{https://doi.org/10.1103/PhysRevD.7.1888}{\emph{Phys. Rev. D} {\bfseries
  7} (1973) 1888}.

\bibitem{Dolan:1973qd}
L.~Dolan and R.~Jackiw, \emph{{Symmetry Behavior at Finite Temperature}},
  \href{https://doi.org/10.1103/PhysRevD.9.3320}{\emph{Phys. Rev. D} {\bfseries
  9} (1974) 3320}.

\bibitem{Nakayama:2008ks}
K.~Nakayama and F.~Takahashi, \emph{{Cosmological Moduli Problem from Thermal
  Effects}}, \href{https://doi.org/10.1016/j.physletb.2008.11.046}{\emph{Phys.
  Lett. B} {\bfseries 670} (2009) 434}
  [\href{https://arxiv.org/abs/0811.0444}{{\ttfamily 0811.0444}}].

\bibitem{Anguelova:2009ht}
L.~Anguelova, V.~Calo and M.~Cicoli, \emph{{LARGE Volume String
  Compactifications at Finite Temperature}},
  \href{https://doi.org/10.1088/1475-7516/2009/10/025}{\emph{JCAP} {\bfseries
  10} (2009) 025} [\href{https://arxiv.org/abs/0904.0051}{{\ttfamily
  0904.0051}}].

\bibitem{Gallego:2020vbe}
D.~Gallego, \emph{{Finite temperature effects in modular cosmology}},
  \href{https://doi.org/10.1088/1475-7516/2020/09/033}{\emph{JCAP} {\bfseries
  09} (2020) 033} [\href{https://arxiv.org/abs/2005.03939}{{\ttfamily
  2005.03939}}].

\bibitem{blumenhagenStatisticsSupersymmetricDbrane2005}
R.~Blumenhagen, F.~Gmeiner, G.~Honecker, D.~Lust and T.~Weigand, \emph{{The
  Statistics of supersymmetric D-brane models}},
  \href{https://doi.org/10.1016/j.nuclphysb.2005.02.005}{\emph{Nucl. Phys. B}
  {\bfseries 713} (2005) 83}
  [\href{https://arxiv.org/abs/hep-th/0411173}{{\ttfamily hep-th/0411173}}].

\bibitem{fontSupersymmetryBreakingDuality1990a}
A.~Font, L.E.~Ibáñez, D.~Lüst and F.~Quevedo, \emph{Supersymmetry breaking
  from duality invariant gaugino condensation},
  \href{https://doi.org/10.1016/0370-2693(90)90665-S}{\emph{Physics Letters B}
  {\bfseries 245} (1990) 401}.

\bibitem{ibanezStringTheoryParticle2012}
L.E.~Ibanez and A.M.~Uranga, \emph{{String theory and particle physics: An
  introduction to string phenomenology}}, Cambridge University Press (2, 2012).

\bibitem{buchmullerSupersymmetricStandardModel2006}
W.~Buchmuller, K.~Hamaguchi, O.~Lebedev and M.~Ratz, \emph{{Supersymmetric
  standard model from the heterotic string}},
  \href{https://doi.org/10.1103/PhysRevLett.96.121602}{\emph{Phys. Rev. Lett.}
  {\bfseries 96} (2006) 121602}
  [\href{https://arxiv.org/abs/hep-ph/0511035}{{\ttfamily hep-ph/0511035}}].

\bibitem{lebedevHeteroticMinilandscapeII2008}
O.~Lebedev, H.P.~Nilles, S.~Ramos-Sanchez, M.~Ratz and P.K.S.~Vaudrevange,
  \emph{{Heterotic mini-landscape. (II). Completing the search for MSSM vacua
  in a Z(6) orbifold}},
  \href{https://doi.org/10.1016/j.physletb.2008.08.054}{\emph{Phys. Lett. B}
  {\bfseries 668} (2008) 331}
  [\href{https://arxiv.org/abs/0807.4384}{{\ttfamily 0807.4384}}].

\bibitem{lebedevHeteroticRoadMSSM2008}
O.~Lebedev, H.P.~Nilles, S.~Raby, S.~Ramos-Sanchez, M.~Ratz, P.K.S.~Vaudrevange
  et~al., \emph{{The Heterotic Road to the MSSM with R parity}},
  \href{https://doi.org/10.1103/PhysRevD.77.046013}{\emph{Phys. Rev. D}
  {\bfseries 77} (2008) 046013}
  [\href{https://arxiv.org/abs/0708.2691}{{\ttfamily 0708.2691}}].

\bibitem{cremadesSUSYQuiversIntersecting2002}
D.~Cremades, L.E.~Ibanez and F.~Marchesano, \emph{{SUSY quivers, intersecting
  branes and the modest hierarchy problem}},
  \href{https://doi.org/10.1088/1126-6708/2002/07/009}{\emph{JHEP} {\bfseries
  07} (2002) 009} [\href{https://arxiv.org/abs/hep-th/0201205}{{\ttfamily
  hep-th/0201205}}].

\bibitem{Blumenhagen:2003jy}
R.~Blumenhagen, D.~Lust and S.~Stieberger, \emph{{Gauge unification in
  supersymmetric intersecting brane worlds}},
  \href{https://doi.org/10.1088/1126-6708/2003/07/036}{\emph{JHEP} {\bfseries
  07} (2003) 036} [\href{https://arxiv.org/abs/hep-th/0305146}{{\ttfamily
  hep-th/0305146}}].

\bibitem{lustScatteringGaugeMatter2004}
D.~Lust, P.~Mayr, R.~Richter and S.~Stieberger, \emph{{Scattering of gauge,
  matter, and moduli fields from intersecting branes}},
  \href{https://doi.org/10.1016/j.nuclphysb.2004.06.052}{\emph{Nucl. Phys. B}
  {\bfseries 696} (2004) 205}
  [\href{https://arxiv.org/abs/hep-th/0404134}{{\ttfamily hep-th/0404134}}].

\bibitem{grimmEffectiveActionCalabi2004}
T.W.~Grimm and J.~Louis, \emph{{The Effective action of N = 1 Calabi-Yau
  orientifolds}},
  \href{https://doi.org/10.1016/j.nuclphysb.2004.08.005}{\emph{Nucl. Phys. B}
  {\bfseries 699} (2004) 387}
  [\href{https://arxiv.org/abs/hep-th/0403067}{{\ttfamily hep-th/0403067}}].

\bibitem{jockersEffectiveActionD7branes2005}
H.~Jockers and J.~Louis, \emph{{The Effective action of D7-branes in N = 1
  Calabi-Yau orientifolds}},
  \href{https://doi.org/10.1016/j.nuclphysb.2004.11.009}{\emph{Nucl. Phys. B}
  {\bfseries 705} (2005) 167}
  [\href{https://arxiv.org/abs/hep-th/0409098}{{\ttfamily hep-th/0409098}}].

\bibitem{fontSUSYbreakingSoftTerms2005}
A.~Font and L.E.~Ibanez, \emph{{SUSY-breaking soft terms in a MSSM magnetized
  D7-brane model}},
  \href{https://doi.org/10.1088/1126-6708/2005/03/040}{\emph{JHEP} {\bfseries
  03} (2005) 040} [\href{https://arxiv.org/abs/hep-th/0412150}{{\ttfamily
  hep-th/0412150}}].

\bibitem{cicoliToricK3fibredCalabiYau2012a}
M.~Cicoli, M.~Kreuzer and C.~Mayrhofer, \emph{{Toric K3-Fibred Calabi-Yau
  Manifolds with del Pezzo Divisors for String Compactifications}},
  \href{https://doi.org/10.1007/JHEP02(2012)002}{\emph{JHEP} {\bfseries 02}
  (2012) 002} [\href{https://arxiv.org/abs/1107.0383}{{\ttfamily 1107.0383}}].

\end{thebibliography}\endgroup
\end{document}